\documentclass[useAMS,usenatbib,usegraphicx]{mn2e}

\usepackage{graphicx,times,epsfig,rotating}
\usepackage{hyperref}

\usepackage{color}

\def\mnras{MNRAS}
\def\aj{AJ}
\def\aap{A\&A}
\def\apj{ApJ}
\def\apjl{ApJ}
\def\apjs{ApJS}
\def\araa{ARA\&A}
\def\pasp{PASP}
\def\nat{Nature}
\def\physrep{Phys.\ Rep.}
\def\aaps{A\&AS}

\newcommand{\kmsend}{\mbox{km s$^{-1}$}}
 \newcommand{\kms}{\mbox{km s$^{-1}$ }}

\newcommand{\hpc}{\mbox{h$^{-1}$pc }}

\newcommand{\msun}{\mbox{M$_{\sun}$ }}
\newcommand{\msunend}{\mbox{M$_{\sun}$}}
\newcommand{\lsun}{\mbox{L$_{\sun}$ }}

\newcommand{\lir}{\mbox{L$_{\rm IR}$}}

\newcommand{\cmthree}{\mbox{cm$^{-3}$}}

\newcommand{\msunyr}{\mbox{M$_{\sun}$ yr$^{-1}$ }}
\newcommand{\msunyrend}{\mbox{M$_{\sun}$ yr$^{-1}$}}
\newcommand{\htwo}{\mbox{H$_2$}}

\newcommand{\z}{\mbox{$z$}}

\newcommand{\zsim}{\mbox{$z\sim$ }}

\newcommand{\bzk}{\mbox{{\it BzK}}}

\newcommand{\xco}{\mbox{$X_{\rm CO}$}}
\newcommand{\xcounits}{\mbox{cm$^{-2}$/K-km s$^{-1}$}}
\newcommand{\qeos}{\mbox{$q_{\rm EOS}$}}

\newcommand{\alphaco}{\mbox{$\alpha_{\rm CO}$}}
\newcommand{\alphacounits}{\mbox{\msun pc$^{-2}$ (K-\kmsend)$^{-1}$}}

\newcommand{\sunrise}{\mbox{\sc sunrise}}
\newcommand{\gadget}{\mbox{\sc gadget-3}}

\newcommand{\starburst}{\mbox{\sc starburst99}}
\newcommand{\mappings}{\mbox{\sc mappingsiii}}
\newcommand{\turtlebeach}{\mbox{\sc turtlebeach}}
\newcommand{\despotic}{\mbox{\sc despotic}}
\newcommand{\astropy}{\mbox{\sc astropy}}
\newcommand{\yt}{\mbox{\sc yt}}
\newcommand{\galform}{\mbox{\sc galform}}

\title[A Theory for CO Excitation in Galaxies]{A
  Theory for the Excitation of CO in Star Forming Galaxies}

\author[Narayanan \& Krumholz]{Desika\, Narayanan$^{1}$\thanks{E-mail:
    dnarayan@haverford.edu} \& Mark R.
  Krumholz$^{2}$\\$^{1}$Department of Physics and Astronomy, Haverford
  College, 370 Lancaster Ave, Haverford, PA 19041\\$^{2}$Department of
  Astronomy and Astrophysics, University of California, Santa Cruz,
  CA, 95064
  }
\begin{document}

\date{Submitted to MNRAS}

\pagerange{\pageref{firstpage}--\pageref{lastpage}} \pubyear{2010}

\maketitle

\label{firstpage}

\begin{abstract}
Observations of molecular gas in high-\z \ star-forming galaxies
typically rely on emission from CO lines arising from states with
rotational quantum numbers $J > 1$. Converting these observations to
an estimate of the CO $J=1- 0$ intensity, and thus inferring \htwo
\ gas masses, requires knowledge of the CO excitation ladder, or
spectral line energy distribution (SLED).  The few available multi-$J$
CO observations of galaxies show a very broad range of SLEDs, even at
fixed galaxy mass and star formation rate, making the conversion to
$J=1- 0$ emission and hence molecular gas mass highly uncertain.
Here, we combine numerical simulations of disk galaxies and galaxy
mergers with molecular line radiative transfer calculations to develop
a model for the physical parameters that drive variations in CO SLEDs
in galaxies.  An essential feature of our model is a fully
self-consistent computation of the molecular gas temperature and
excitation structure. We find that, while the shape of the SLED is
ultimately determined by difficult-to-observe quantities such as the
gas density, temperature, and optical depth distributions, all of these
quantities are well-correlated with the galaxy's mean star formation
rate surface density ($\Sigma_{\rm SFR}$), which is observable. We use
this result to develop a model for the CO SLED in terms of
$\Sigma_{\rm SFR}$, and show that this model quantitatively reproduces
the SLEDs of galaxies over a dynamic range of $\sim 200$ in SFR
surface density, at redshifts from $z=0-6$.  This model should make it
possible to significantly reduce the uncertainty in deducing molecular
gas masses from observations of high-$J$ CO emission.  
\end{abstract}
\begin{keywords}
ISM: clouds -- ISM: molecules -- galaxies: ISM -- galaxies: interactions
-- galaxies: star formation -- galaxies: starburst
\end{keywords}

\section{Introduction}
\label{section:introduction}

Stars form in giant clouds in galaxies, comprised principally of
molecular hydrogen, \htwo.  Due to its low mass, \htwo \ requires
temperatures of $\sim 500 $ K to excite the first quadrapole line,
while GMCs in quiescent galaxies have typical temperatures of $\sim
10$ K \citep[e.g.][]{fuk10, dob13,kru14a}.  The relatively low abundance of emitting
\htwo \ in molecular clouds has driven the use of the ground state
rotational transition of carbon monoxide, $^{12}$CO (hereafter, CO),
as the most commonly observed tracer molecule of \htwo.  With a
typical abundance of $\sim 1.5 \times 10^{-4} \times \htwo$ at solar
metallicity, CO is the second most abundant molecule in clouds
\citep{lee96}.  Via a (highly-debated) conversion factor from CO
luminosities to \htwo \ gas masses, observations of the $J=1- 0$
transition of CO provide measurements of molecular gas masses,
surface densities, and gas fractions in galaxies at both low and
high-redshifts.

CO can be excited to higher rotational states via a combination of
collisions with \htwo \ and He, as well as through radiative
absorptions.  When the CO is thermalized, then level populations can
be described by Boltzmann statistics, and the intensity at a given
level follows the Planck function at that frequency.

While the conversion factor from CO luminosities to \htwo \ gas masses
is based solely on the $J=1- 0$ transition, directly detecting the ground
state of CO at high-\z \ is challenging.  Only a handful of facilities
operate at long enough wavelengths to detect the redshifted 2.6 mm CO
$J=1- 0$ line, and consequently the bulk of CO detections at
high-redshift are limited to high-$J$ transitions \citep[see the
  compendia in recent reviews by][]{sol05,car13}. Thus, to arrive at
an \htwo \ gas mass, one must first down-convert from the observed
high-$J$ CO line intensity to a $J=1- 0$ intensity before applying a
CO-\htwo \ conversion factor.  This requires some assumption of the CO
excitation ladder, or as it is often monikered in the literature, the
Spectral Line Energy Distribution (SLED).

Typically, line ratios between a high-$J$ line and the $J=1- 0$
line are assumed based on either an assumption of thermalized level
populations, or by comparing to a similar galaxy (selected either via
similar methods, or having similar inferred physical properties such
as SFR) \citep[e.g.][]{gea12,car13}.  However, this method is
uncertain at the order of magnitude level, even for lines of modest
excitation such as $J=3- 2$.

\begin{figure}
\includegraphics[scale=0.45]{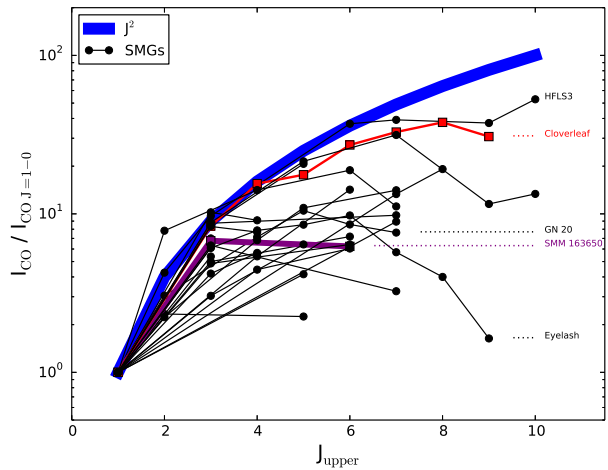}
\caption{CO Spectral Line Energy Distributions (SLEDs) for all known high-\z
  \ Submillimeter Galaxies with a CO $J=1- 0$ detection.  A significant
  diversity in CO excitations exists, even for a given class of high-\z
  \ galaxies, and it is evident that no universal line ratios are
  applicable.  For a given line, roughly an order of magnitude
  uncertainty exists in the conversion ratio from high-$J$ lines to the
  ground state. The blue line denotes $J^2$, the scaling of
  intensities expected if the lines are all in the Rayleigh-Jeans
  limit and in local thermodynamic equilibrium (LTE).  The red and purple lines denote the Cloverleaf
  quasar and SMG SMM 163650 - two galaxies that have similar star
  formation rates, but starkly different
  SLEDs.\label{figure:cosled_observational}}
\end{figure}

As an anecdotal example, consider the case of high-\z
\ submillimeter galaxies (SMGs).  These 850 $\micron$-selected
galaxies typically live from $z=2-4$, have star formation
rates well in excess of 500 \msunyrend, and are among the most
luminous galaxies in the Universe \citep[see the reviews by][]{bla02,cas14a}.  Originally, it was assumed that the gas was
likely thermalized in these systems, given their extreme star
formation rates (and likely extreme ISM densities and temperatures)
\citep[e.g.][]{gre05,cop08,tac08}.  However, subsequent CO
$J=1- 0$ observations
of a small sample of SMGs with higher $J$ detections by \citet{hai06}
and \citet{har10} showed that a number of SMGs exhibit subthermal
level populations, even at the modest excitation level of $J=3$.
Since then, it has become increasingly clear that SMGs are a diverse
population
\citep[e.g.][]{bot13,hay11,hay12a,hay13b,hay13a,hod13,kar13}, with a
large range in potential CO excitation ladders.  To quantify this, in
Figure~\ref{figure:cosled_observational}, we have compiled all of the
CO SLEDs published to date for all high-\z \ SMGs with a CO $J=1- 0$
detection.  The diversity in the observed SLEDs is striking, even for
one of the most extreme star forming galaxy populations known.  It is
clear that no ``template'' SLED exists for a given population.

Nor can one do much better simply by comparing an observed galaxy to
another galaxy with a known SLED and a similar luminosity or SFR.  For
example, turning again to Figure~\ref{figure:cosled_observational}, we
highlight the SLED of a single quasar host galaxy denoted by the red
line (the Cloverleaf quasar), and compare it to SMG SMM 163650
(purple).  Both galaxies have estimated star formation rates of
roughly $\sim 1000 \ \msunyrend$ \citep{ivi11,sol03,wei03}, but they
have dramatically different SLEDs.  For the $J=6- 5$ line, there is
roughly a factor of five difference in the SLEDs, and the discrepancy
likely grows at even higher levels.  It is clear from this example
that extrapolating from high $J$ lines to $J=1- 0$ using templates
based on galaxies with similar global SFRs is a problematic
approach.

 The consequences of uncertainty in the excitation of CO are
 widespread in the astrophysics of star formation and galaxy
 formation.  For example, observed estimates of the Kennicutt-Schmidt
 star formation relation at high redshift, an often-used observational
 test bed for theories of star formation in environments of extreme
 pressure \citep[e.g.][]{ost11,kru12a,fau13}, are dependent on assumed
 CO line ratios.  Similarly, baryonic gas fractions of
 galaxies at high-\z, an observable diagnostic of the equilibrium
 established by gaseous inflows from the intergalactic medium, gas
 consumption by star formation, and gas removal by winds, depend
 critically on observations of high-$J$ CO lines, and their conversion
 to the $J=1- 0$ line
 \citep[e.g.][]{tac10,dav12,kru11e,bot12,nar12c,sai13,for13,hop13b,hop13c}.
 In the near future, CO deep fields will measure the CO luminosity
 function over a wide range of redshifts, and our ability to convert
 the observed emission into an estimate of $\Omega_{\rm H_2}$, the
 cosmic baryon budget tied up in H$_2$ molecules, will depend on our
 understanding of the systematic evolution of CO SLEDs.

While some theoretical work has been devoted to modeling CO SLEDs from
galaxies
\citep{nar08a,nar08d,nar09,arm12,lag12,mun13b,mun13,pop13b,pop13},
these models have by and large focused on utilizing synthetic CO
excitation patterns to validate their underlying galaxy formation
model.  We lack a first-principles physical model for the origin of the shape of
SLEDs in galaxies, and the relationship of this shape to observables.  The goal of
this paper is to provide one. The principle idea is that CO
excitation is dependent on the gas temperatures, densities and optical
depths in the molecular ISM.  Utilizing a detailed physical model for
the thermal and physical structure of the star-forming ISM in both
isolated disk galaxies, and starbursting galaxy mergers, we
investigate what sets these quantities in varying physical
environments.  We use this information to provide a
parameterization of CO SLEDs in terms of the star formation rate
surface density ($\Sigma_{\rm SFR}$), and show that this model
compares very favorably with
observed SLEDs for galaxies across a range of physical conditions and
redshifts.

This paper is organized as follows. In \S~\ref{section:methods}, we
describe our numerical modeling techniques, including details
regarding our hydrodynamic models, radiative transfer techniques, and
ISM specification.  In \S~\ref{section:results}, we present our main
results, including the principle drivers of SLED shapes, as well as a
parameterized model for observed SLEDs in terms of the galaxy
$\Sigma_{\rm SFR}$.  In \S~\ref{section:discussion}, we provide
discussion, and we conclude in \S~\ref{section:conclusions}.

\begin{table*}
\label{table:ICs}
\centering
\begin{minipage}{100mm}
\caption{Galaxy evolution simulation parameters. Column 1 refers to
  the model name. Galaxies with the word 'iso' in them are isolated
  disks, and the rest of the models are 1:1 major mergers. Column 2
  is the baryonic mass in \msunend.  Columns 3-6 refer to the orbit of
  a merger (with ``N/A'' again referring to isolated discs).  Column 7
  is the initial baryonic gas fraction of the galaxy, and Column 8 refers
  to the redshift of the simulation. }
\begin{tabular}{@{}cccccccc@{}}
\hline Model  & $M_{\rm bar}$&  $\theta_1$&$\phi_1$ & $\theta_2$ & $\phi_2$ &$f_g$&z\\ 
&\msun&&&&&&\\ 
 1 &2 &3 &4 &5 &6 &7 &8\\ \hline \hline\\
z3isob6 &$3.8 \times 10^{11}$&N/A&N/A&N/A&N/A&0.8&3\\
z3isob5 &$1.0 \times 10^{11}$&N/A&N/A&N/A&N/A&0.8&3\\
z3isob4 &$3.5 \times 10^{10}$&N/A&N/A&N/A&N/A&0.8&3\\
z3b5e   &$2.0 \times 10^{11}$&30&60&-30&45&0.8&3\\
z3b5i   &$2.0 \times 10^{11}$&0&0&71&30&0.8&3\\
z3b5j   &$2.0 \times 10^{11}$&-109&90&71&90&0.8&3\\
z3b5l   &$2.0 \times 10^{11}$&-109&30&180&0&0.8&3\\
z0isod5 &$4.5 \times 10^{11}$&N/A&N/A&N/A&N/A&0.4&0\\
z0isod4 &$1.6 \times 10^{11}$&N/A&N/A&N/A&N/A&0.4&0\\
z0isod3 &$5.6 \times 10^{10}$&N/A&N/A&N/A&N/A&0.4&0\\
z0d4e   &$3.1 \times 10^{11}$&30&60&-30&45&0.4&0\\
z0d4h   &$3.1 \times 10^{11}$&0&0&0&0&0.4&0\\
z0d4i   &$3.1 \times 10^{11}$&0&0&71&30&0.4&0\\
z0d4j   &$3.1 \times 10^{11}$&-109&90&71&90&0.4&0\\
z0d4k   &$3.1 \times 10^{11}$&-109&30&71&-30&0.4&0\\
z0d4l   &$3.1 \times 10^{11}$&-109&30&180&0&0.4&0\\
z0d4n   &$3.1 \times 10^{11}$&-109&-30&71&30&0.4&0\\
z0d4o   &$3.1 \times 10^{11}$&-109&30&71&-30&0.4&0\\
z0d4p   &$3.1 \times 10^{11}$&-109&30&180&0&0.4&0\\

\hline
\end{tabular}
\end{minipage}
\end{table*}

\section{Numerical Modeling}
\label{section:methods}

\subsection{Overview of Models}

The basic strategy that we employ is as follows, and is nearly
identical to that employed by \citet{nar11b, nar12a}.  We first
simulate the evolution of idealized disk galaxies and galaxy mergers
over a range of masses, merger orbits, and halo virial properties
utilizing the smoothed particle hydrodynamics (SPH) code, \gadget
\ \citep{spr03a,spr05b,spr05a}.  The goal of the idealized simulations
is not to provide a simulated analog of any particular observed
galaxy, but rather to simulate a large range of physical conditions.
As we will show, our simulations encapsulate a large range of
molecular cloud densities, temperatures, velocity dispersions, and CO
optical depths.

We project the physical conditions of the SPH particles onto an
adaptive mesh, and calculate the physical and chemical properties of
the molecular ISM in post-processing.  In particular, we determine the
temperatures, mean cloud densities, velocity dispersions, \htwo-H~\textsc{i}
balance, and fraction of carbon locked in CO.  Determining the
thermodynamic structure of the ISM involves balancing heating
by cosmic rays and the grain photoelectric effect, energy
exchange with dust, and CO and C~\textsc{ii} line cooling.  To calculate the
line cooling rates, we employ the public code to Derive the Energetics
and SPectra of Optically Thick Interstellar Clouds
\citep[\despotic;][]{kru13}, built on the escape probability
formalism.  The radiative transfer calculations determine the level
populations via a balance of radiative absorptions, stimulated
emission, spontaneous emission, and collisions with \htwo \ and He.
With these level populations in hand, we determine full CO rotational
ladder for each cloud within our model galaxies.  The observed SLED
from the galaxy is just the combination of the CO line intensities at
each level from each cloud in a given system.

At this point, the general reader should have enough background to
continue to the Results (\S~\ref{section:results}) if they so wish.
For those interested in the technical aspects of the simulation
methodology, we elaborate further in the remainder of this section.

\subsection{Hydrodynamic Simulations of Galaxies in Evolution}
\label{section:sph}

The main purpose of the SPH simulations of galaxies in evolution is to
generate a large number of realizations of galaxy models with a
diverse range of physical properties.  To this end, we simulate the
hydrodynamic evolution of both idealized disks, as well as mergers
between these disks.  Here, we summarize the features of the
simulations most pertinent to this paper, and we refer the reader to
\citet{spr02,spr03a} and \citet{spr05a} for a more detailed
description of the underlying algorithms for the hydrodynamic code
employed, \gadget.

The galaxies are exponential disks, and are initialized following the
\citet{mo98} formalism.  They reside in live dark halos, initialized
with a \citet{her90} profile.  We simulate galaxies that have halo
concentrations and virial radii scaled for $z=0$ and $z=3$, following
the relations of \citet{bul01} and \citet{rob06a}.  The latter
redshift choice is motivated by the desire to include a few extreme
starburst galaxies in our simulation sample.  The gravitational
softening length for baryons is 100 \hpc, and 200 \hpc for dark
matter.

The gas is initialized as primoridal, with all metals forming as the
simulation evolves.  The ISM is modeled as multi-phase, with clouds
embedded in a pressure-confining hotter phase \citep[e.g.][]{mck77}.
In practice, this is implemented in the simulations via hybrid SPH
particles.  Cold clouds grow via radiative cooling of the hotter
phase, while supernovae can heat the cold gas.  Stars form in these
clouds following a volumetric \citet{sch59} relation, with index 1.5
\citep{ken98a,cox06b,nar11a,kru12a}.  We note that in \citet{nar11b}, we
explored the effects of varying the Kennicutt-Schmidt star formation
law index, and found that the ISM thermodynamic properties were
largely unchanged so long as the index is $\geq 1$.  In either case,
observations \cite[e.g.][]{big08}, as well as theoretical work
\cite[e.g.][]{kru09b,ost11} suggest that the index may be
superlinear in high-surface density starburst environments. 

 The pressurization of the ISM via supernovae is implemented via an
 effective equation of state \citep{spr05a}.  In the models scaled for
 present epoch, we assume a modest pressurization (\qeos = 0.25),
 while for the high-\z \ galaxies, we assume a more extreme
 pressurization (\qeos=1).  Tests by \citet{nar11b} show that the
 extracted thermodynamic properties of the simulations are relatively
 insensitive to the choice of EOS.  Still, we are intentional with
 these choices.  The high-\z \ galaxies would suffer from runaway
 gravitational fragmentation, preventing starbursts, without sufficient
 pressurization.  Because the simulations are not cosmological, and do
 not include a hot halo \citep[e.g.][]{mos11a,mos12} or resolved
 feedback \citep{kru10,kru12b,kru13b,hop13b,hop13}, the early phases of
 evolution would quickly extinguish the available gas supply without
 sufficient pressurization.  Despite this pressurization, large scale
 kpc-sized clumps do still form from dynamical instabilities in
 high-\z \ systems, similar to observations
 \citep{elm08,elm09b,elm09a} and simulations
 \citep[e.g.][]{dek09,cev10}.

For the galaxy mergers, we simulate a range of arbitrarily chosen
orbits.  In Table~\ref{table:ICs}, we outline the basic physical
properties of the galaxies simulated here.

\subsection{The Physical and Chemical Properties of the Molecular ISM}
\label{section:ism}

We determine the physical and chemical properties of the molecular
interstellar medium in postprocessing.  We project the physical
properties of the SPH particles onto an adaptive mesh, utilizing an
octree memory structure.  Practically, this is done through the dust
radiative transfer code \sunrise
\ \citep{jon06a,jon06b,jon10a,jon10b}.  The base mesh is 5$^3$ cells,
spanning a 200 kpc box.  The cells recursively refine into octs based
on the refinement criteria that the relative density variations of
metals ($\sigma_{\rho_m}/\langle \rho_m \rangle$) should be less than
0.1.  The smallest cell size in the refined mesh are $\sim 70$ pc
across, reflecting the minimum cell size necessary for a converged
dust temprature \citep{nar10a}.

We assume that all of the neutral gas in each cell is locked into
spherical, isothermal clouds of constant density.  The surface density
is calculated directly from the gas mass and cell size.  For cells
that are very large, we consider the cloud unresolved, and adopt a
subresolution surface density of $\Sigma_{\rm cloud} = 85 \ \msun$
pc$^{-2}$, comparable to observed values of local GMCs
\citep[e.g.][]{sol87,bol08,dob13}.  The bulk of the \htwo \ gas mass in our
model galaxies, though, come from resolved GMCs as the large cells in
the adaptive mesh (that force the imposition of the threshold cloud
surface density) correspond to low density regions.

The \htwo \ fraction within the neutral ISM is determined utilizing the
analytic formalism of \citet{kru08,kru09a,kru09b}.  In short, this
model balances the photodissociation of \htwo \ molecules by
Lyman-Werner band photons against the growth of molecules on grains.
Formally this equilibrium model for the ISM molecular fraction is given by:
\begin{equation}
\label{eq:kmt}
f_{\rm H2} \approx 1 - \frac{3}{4}\frac{s}{1+0.25s}
\end{equation}
for $s<2$ and $f_{\rm H2} = 0$ for $s\geq 2$, where $s = {\rm ln}
(1+0.6\chi + 0.01\chi^2)/(0.6\tau_{\rm c})$, $\chi =
0.76(1+3.1Z'^{0.365})$, and $\tau_{\rm c} = 0.066 \Sigma_{\rm
  cloud}/(\msun {\rm pc^{-2}})\times Z'$. Here $Z'$ is the metallicity
divided by the solar metallicity.  We motivate the usage of the
\citet{kru09b} model for \htwo \ formation and destruction for two
reasons.  First, observations by \citet{fum10} and \citet{won13} of dwarf galaxies
suggest that the adopted method in Equation~\ref{eq:kmt} may fare
better at low metallicities than observed empirical pressure-based
relations \citep[e.g.][]{bli06}.  Second, \citet{kru11b} find that
Equation~\ref{eq:kmt} compares well to full numerical time-dependent
chemical reaction networks and radiative transfer models for \htwo
\ formation and destruction \citep{pel06,gne09,gne10,chr12b,chr12,mun13c} above
metallicities of $\sim 0.1 Z_\odot$.  The mass-weighted metallicity in
our star-forming galaxies are always above this threshold.

In order to account for the turbulent compression of gas, we scale the
volumetric densities of the clouds by a factor $e^{\sigma_\rho^2/2}$,
where $\sigma_{\rho}^2$ is approximated by:
\begin{equation}
\label{eq:turbulentcompression}
\sigma_\rho^2 \approx {\rm ln}(1+3M_{\rm 1D}^2/4)
\end{equation}
based on insight from numerical simulations \citep{ost01,pad02,lem08}.
Here, $M_{\rm 1D}$ is the one-dimensional Mach number of the
turbulence.  In order to calculate the Mach number, we assume the
temperature of the GMC is 10 K.  This represents a minor inconsistency
in our modeling, as for the remainder of the gas specification, we
utilize a fully radiative model to determine the gas thermal
properties (\S~\ref{section:thermodynamics}).\footnote{We use this
simplification to avoid having to iterate the Mach number, gas
temperature, dust temperature, and level populations to convergence
simultaneously, as this would be prohibitively expensive. The current
calculation, using a fixed Mach number while iterating the gas and
dust temperatures and level populations to convergence, is already
quite expensive. The effect of assuming a fixed Mach number is fairly
small in any event. The Mach number enters the calculation
\textit{only} through its effects on the median gas density used in
calculations of the collisional excitation rates, so the main effect
of a fully self-consistent treatment would be to lower the Mach number
and thus the median gas density in the most vigorously star-forming
galaxies where the molecular gas temperature exceeds 10 K. However, in
these galaxies the density is so high that deviations from LTE are
generally not very important, and would remain unimportant even with
lower Mach numbers.}

The 1D velocity dispersion of in the cloud is determined by:
\begin{equation}
\sigma = {\rm max}( \sigma_{\rm cell},\sigma_{\rm vir})
\end{equation}
where $\sigma_{\rm cell}$ is the mean square sum of the subgrid
turbulent velocity dispersion within the GMC and the resolved
nonthermal velocity dispersion.  The external pressure from the hot
ISM determines the subgrid turbulent velocity dispersion using
$\sigma^2=P/\rho_{cell}$, with a ceiling of 10 \kms \ imposed.  This
ceiling value comes from average values seen in turbulent feedback
simulations \citep{dib06,jou09,ost11}.  The resolved non-thermal
component is determined via the mean of the standard deviation of the
velocities of the nearest neighbour cells in the $\hat{x}$, $\hat{y}$
and $\hat{z}$ directions.  In cases where the GMC is unresolved, a
floor velocity dispersion, $\sigma_{\rm vir}$ is set assuming the GMC
is in virial balance with virial parameter $\alpha_{\rm vir} = 1$.

\subsection{Thermodynamics and Radiative Transfer}
\label{section:thermodynamics}

A key part to calculating the observed SLED is the determination of
the gas temperature in the model GMCs.  The model is rooted in
algorithms presented by \citet{gol01} and \citet{kru11a}, and tested
extensively in models presented by \citet{nar11b,nar12a}.
\citet{mun13b,mun13} also use a similar approach. All our calculations
here make use of the \textsc{despotic} package \citep{kru13}, and
our choices of physical parameters match the defaults in
\textsc{despotic} unless stated otherwise.

The temperature of the \htwo \ gas is set by a balance of
heating and cooling processes in the gas, heating and cooling in the
dust, and energy exchange between the gas and dust\footnote{In
  \citet{nar11b}, we explored models that included the effects of
  turbulent heating on molecular clouds via both adiabatic compression
  as well as viscous dissipation.  These tests found only modest
  differences between our fiducial model and models that included
  turbulent heating.}.  The principal gas heating processes are via
cosmic rays and the grain photoelectric effect, and energy exchange
with dust.  The principle coolant is CO when the bulk of the carbon is
in molecular form, else C~\textsc{ii} when in atomic form. The dominant heating
mechanism for dust is absorption of radiation, while the cooling
occurs via thermal emission.  Formally, the equations of thermal
balance for gas and dust are:
\begin{eqnarray}
\label{eq:gastemp}
\Gamma_{\rm pe} + \Gamma_{\rm CR} - \Lambda_{\rm line} + \Psi_{\rm gd} = 0\\
\label{eq:dusttemp}
\Gamma_{\rm dust} - \Lambda_{\rm dust} - \Psi_{\rm gd} = 0  
\end{eqnarray}
where $\Gamma$ terms represent heating rates, $\Lambda$ terms
are cooling rates, $\Psi_{\rm gd}$ is the dust-gas energy exchange rate,
which may be in either direction depending on their relative temperatures.
We solve these equations by simultaneously iterating on the gas and dust temperatures.

The grain photoelectric effect heats the gas via FUV photons ejecting
electrons from dust grains.  Because FUV photons likely suffer heavy
extinction in clouds, we attenuate the grain photoelectric heating
rate by half the mean extinction in the clouds in order to account for
reduced heating rates in cloud interiors:
\begin{equation}
\Gamma_{\rm pe} = 4 \times 10^{-26}G_0'e^{-N_{\rm H}\sigma_{\rm d}/2} {\rm erg \ s^{-1}}
\end{equation}
where $G_0'$ is the FUV intensity relative to the Solar neighborhood,
and $\sigma_{\rm d}$ is the dust cross section per H atom to UV
photons (given by the value $\sigma_d = 1 \times 10^{-21}$ cm$^{-2}$).
We assume that $G_0'$ scales linearly with the galaxy SFR, normalized
by a Milky Way SFR of 2 \msunyr \ \citep{rob10}.

The cosmic ray heating rate is driven by heated electrons freed by
cosmic-ray induced ionizations, and is given by:
\begin{equation}
\Gamma_{\rm CR} = \zeta' q_{\rm CR} \ {\rm s^{-1}}
\end{equation}
where $\zeta'$ is the cosmic ray primary ionization rate (assumed to be 2
$\times 10^{-17} Z' {\rm s}^{-1}$), and $q_{\rm CR}$ is the thermal
energy increase per cosmic ray ionization.  The values for $q_{\rm
  CR}$ are given in Appendix B1 of \citet{kru13}. As with $G_0'$
before, we assume the cosmic ray ionization rate scales linearly with
the SFR of the galaxy.

The carbon abundance is set to $X_c = 1.5 \times 10^{-4} Z'$
\citep{lee96}, where the fraction of carbon locked into CO molecules
is approximated by the semi-analytic model of \citet{wol10}:
\begin{equation}
\label{eq:abundance}
f_{\rm CO} = f_{\rm H2} \times e^{-4(0.53-0.045 {\rm
    ln}\frac{G_0'}{n_{\rm H}/{\rm cm^{-3}}}-0.097 {\rm ln}Z')/A_{\rm v}}
\end{equation}
This model agrees well with the fully numerical calculations of
\citet{glo11}.  When $f_{\rm CO}$ is above 50\%, the cooling is
assumed to happen via CO line cooling; otherwise, we assume that C~\textsc{ii}
dominates the carbon-based line cooling.  The cooling rates are
determined by full radiative transfer calculations, which also determine
the CO line luminosity for each GMC in our model galaxies, as
described below.

The dust cooling rate is: 
\begin{equation}
\Lambda_{\rm dust} = \sigma_d (T_{\rm d})\mu_{\rm H}caT_{\rm d}^4
\end{equation}
where the bulk of the dust heating happens via IR radiation in the
centers of star-forming clouds. The mean dust cross section per H
nucleus, $\sigma_d$, varies with the dust temperature, $T_d$ as a
powerlaw: $\sigma_d = \sigma_{d,10} \times (T_d/10)^2$, and
$\sigma_{d,10} = 2 \times 10^{-25}$ cm$^{2}$ H$^{-1}$.

\begin{figure}
\includegraphics[scale=0.45]{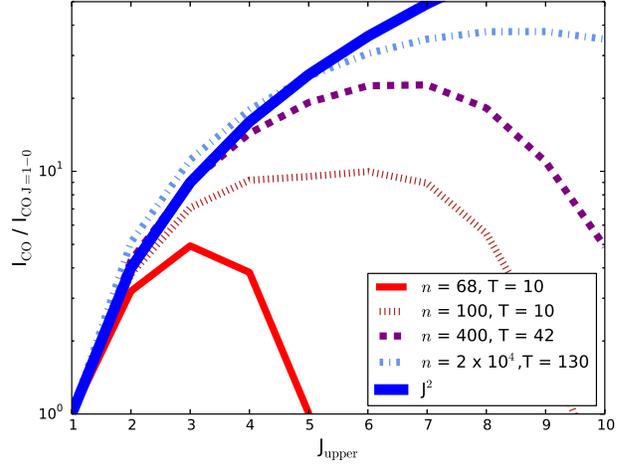}
\caption{Model CO SLEDs for snapshots in a model galaxy merger with
  differing gas temperatures and densities.  The blue line shows $J^2$
  scaling, the expected scaling of intensities for levels in LTE and
  in the Rayleigh-Jeans limit.  As densities and temperatures rise, so
  does the SLED.  It is important to note that an individual large velocity gradient
  or escape probability
  solution with the quoted temperatures and densities may not give the
  same simulated SLED as shown here for our model galaxies. Our model
  galaxy SLEDs result from the the
  superposition of numerous clouds of varying density and temperature. 
  \label{figure:cosled_physical}}
\end{figure}

\begin{figure}
\includegraphics[scale=0.45]{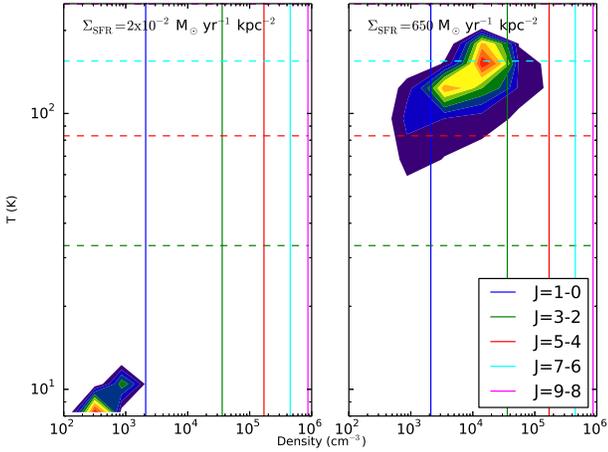}
\caption{ Distribution of GMCs in density-temperature phase space 
  for two example simulated galaxies, one with low $\Sigma_{\rm SFR}$ (left
  panel) and one with high $\Sigma_{\rm SFR}$ (right panel). The main result 
  from this figure is that gas in starburst
  environments is significantly warmer and denser than in quiescent
  environments.  The left panel shows a galaxy with $\Sigma_{\rm SFR}
  \approx 2 \times 10^{-2} \msun$ yr$^{-1}$ kpc$^{-2}$
  (i.e. comparable to nearby quiescent disk galaxies), while the right
  shows a starburst with $\Sigma_{\rm SFR} \approx 650 \ \msun$
  yr$^{-1}$ kpc$^{-2}$ (i.e. comparable to the nuclear regions of
  local ULIRGs and high-\z \ submillimeter galaxies).  The innermost
  contour shows the region in the $n-T$ plane where the \htwo \ mass
  density exceeds $5.5 \times 10^7$($1\times10^9$) \msun dex$^{-2}$ in
  the left (right) panel, and subsequent contours are spaced evenly in
  increments of $5.5 \times 10^6$($1\times10^8$) \msun dex$^{-2}$.
  The horizontal dashed lines show $E_{\rm upper}/k$ for transitions CO
  $J=1- 0$ through $9-8$ (alternating), and the vertical solid lines
  show the critical density, $n_{\rm crit}$ for the same transitions
  at an assumed temperature of $T=100$K, and are taken from the
  \citet{car13} review.  We note that critical densities depend on
  temperature (via collisional rates), though only
  weakly.  \label{figure:phase_space}}
\end{figure}

The dust heating rate is given by:
\begin{equation}
\Gamma_{\rm dust} = \kappa (T_{\rm rad}) \mu_{\rm H}caT_{\rm rad}^4
\end{equation}
We determine the input radiation temperature $T_{\rm rad}$ via
\sunrise \ 3D polychromatic dust radiative transfer calculations
\citep{jon06a,jon10a,jon10b}.  The sources of radiation for the
\sunrise \ input are stellar clusters and accreting black holes
\citep[though the latter have minimal impact on the general thermal
  structure;][]{nar11b}.  The stellar clusters emit a \starburst
\ spectrum, where the stellar ages, metallicites and positions are
known from the \gadget \ simulations.  For the isolated disk galaxies,
young ($< 10$ Myr) stellar clusters are surrounded by H~\textsc{ii} regions and
photodissociation regions (PDRs) whose SEDs are calculated utilizing
1D \mappings \ photoionization modeling \citep{gro04,gro08}.  The
time-averaged covering fraction of PDRs is a free parameter, and we
assume a covering lifetime of 3 Myr as motivated by model tests
presented in \citet{jon10a} who calibrated their models based on a
comparison to SINGS galaxies \citep{ken03}.  In any case,
\citet{nar11b} showed tests that this subresolution parameter choice
has minimal impact on the thermal structure of the ISM.  For galaxy
mergers, we abandon this model: the stellar density in these scenarios
can be high enough that superstellar clusters ($>10^8$ \msun) can
form.  In this case, the \mappings \ photoionization tables saturate
and we assume the cold ISM is a uniform medium with volume filling
factor of unity, motivated in part by observations of nearby ULIRGs
\citep{dow98,sak99}.  Tests by \citet{nar10b,nar11b,hay10} and
\citet{hay11} show that in starburst environments, this choice has
relatively minimal impact on the derived ISM physical properties
\citep[see the Appendix of][]{nar11b}.  The dust mass in the diffuse
ISM is calculated assuming a dust to metals ratio of 0.4
\citep{dwe98,vla98,cal08}, and has \citet{wei01} $R_{\rm V} = 3.15$
opacities.  Utilizing the \citet{juv05} iterative methodology,
\sunrise \ returns the converged radiation field.  We then calculate
the dust temperature in each cell by iterating equations 6-8 of
\citet{jon10b} utilizing a Newton-Raphson scheme as in \citet{nar11b}.
This dust temperature is used as the input $T_{\rm rad}$ in the
\despotic \ calculations.

Finally, the dust and gas exchange energy at a rate
\begin{equation}
\Psi_{\rm gd} = \alpha_{\rm gd}n_{\rm H}T_{\rm g}^{1/2}(T_{\rm d}-T_{\rm g})
\end{equation}
where the thermal gas-dust coupling coefficient is $\alpha_{\rm gd} = 3.2
\times 10^{-34} Z'$ erg cm$^3$ K$^{-3/2}$ for \htwo, and $\alpha_{\rm
  gd } = 1 \times 10^{-33} Z'$ erg cm$^{3}$ K$^{-3/2}$ for H~\textsc{i}
\citep{kru13}.

The CO line cooling rates, and CO line luminosities are calculated
from every GMC in our model galaxies using the escape probability
algorithms built into \despotic.  The CO intensities from a GMC are
set by the level populations.  The source function at a frequency
$\nu$, $S_{\nu}$ for an upper transition to lower $u- l$
is governed by
\begin{equation}
\label{eq:sourcefunction}
S_{\nu}=\frac{n_{u}A_{ul}}{(n_{l}B_{lu}-n_{u}B_{ul})}
\end{equation}
where $A_{ul}, B_{lu}$ and $B_{ul}$ are the Einstein coefficients for
spontaneous emission, absorption, and stimulated emission,
respectively, and $n$ are the absolute level populations.  

The molecular levels are determined by balancing the excitation and
deexcitation processes of the CO rotational lines.  The relevant
processes are collisions with \htwo \ and He, stimulated emission,
absorption, and spontaneous emission.  Formally, these are calculated
via the rate equations
\begin{eqnarray}
\label{eq:kmt_stateq}
\sum_l(C_{lu} + \beta_{lu}A_{lu})f_l =
\left[\sum_u(C_{ul}+\beta_{ul}A_{ul})\right]f_u\\
\sum_if_i = 1
\end{eqnarray}
where $C$ are the collisional rates, $f$ the fractional level
populations, and $\beta_{ul}$ is the escape probability for transition
$u- l$.  The rate equations can be rearranged as an
eigenvalue problem, and solved accordingly.  The Einstein coefficients
and collisional rates are taken from the publicly available {\it Leiden
  Atomic and Molecular Database} \citep{sch05}. In particular, the CO data are
  originally from \citet{yan10b}.  

For a homogeneous, static, spherical cloud, \citet{dra11} shows that
the escape probability, averaged over the line profile and cloud
volume can be approximated by
\begin{equation}
\label{eq:kmt_beta}
\beta_{ul} \approx \frac{1}{1+\frac{3}{8}\tau_{ul}},
\end{equation}
where the optical depth is given by
\begin{equation}
\label{eq:kmt_tau}
\tau_{ul} =
\frac{g_u}{g_l}\frac{3A_{ul}\lambda_{ul}^3}{16(2\pi)^{3/2}\sigma}QN_{\rm
  H2}f_l\left(1-\frac{f_ug_l}{f_lg_u}\right).
\end{equation}
Here $Q$ is the abundance of CO with respect to \htwo, $g_l$ and
$g_u$ are the statistical weights of the levels, $N_{\rm H_2}$ is the
column density of \htwo \ through the cloud, $\lambda_{ul}$ is
the wavelength of the transition, and $\sigma$ is the velocity
dispersion in the cloud.  


\textsc{Despotic} simultaneously solves Equations \ref{eq:kmt_stateq}-\ref{eq:kmt_tau}
with Equations \ref{eq:gastemp} and \ref{eq:dusttemp} using an iteration
process. We refer readers to \citet{kru13} for details. We perform this
procedure for every GMC in every model snapshot of every galaxy
evolution simulation in Table~\ref{table:ICs}. The result is a self-consistent
set of gas and dust temperatures, and CO line luminosities, for every
simulation snapshot. These form the basis for our analysis in the following
section.

\section{Results}
\label{section:results}

  The CO SLED from galaxies represents the excitation of CO, typically
  normalized to the ground state.  As the CO rotational levels are
  increasingly populated, the intensity
  (c.f. Equation~\ref{eq:sourcefunction}) increases, and the SLED
  rises.  When the levels approach local thermodynamic equilibrium
  (LTE), the excitation at those levels saturates.  This is visible in
  the observational data shown in
  Figure~\ref{figure:cosled_observational}, where we showed the SLEDs
  for all SMGs with CO $J=1- 0$ detections.  The Eyelash galaxy has
  modest excitation, with the SLED turning over at relatively low
  $J$-levels.  By comparison, HFLS3, an extreme starburst at \zsim 6
  \citep{rie13}, has levels saturated at $J\sim 7$, with only a weak
  turnover at higher lying lines.
  
  Our simulated galaxies reproduce, at least qualitatively, this range
  of behaviors. We illustrate this in
  Figure~\ref{figure:cosled_physical}, which shows our
  theoretically-computed SLEDs for a sample of snapshots from model
  galaxy merger z0d4e. Each snapshot plotted has a different CO
  mass-weighted mean gas density and temperature.  With increasingly
  extreme physical conditions, the populations in higher $J$ levels
  rise, and the SLED rises accordingly.  Eventually the intensity from
  a given level saturates as it approaches thermalization.
  
  In this Section we more thoroughly explore the behavior of the SLEDs
  in our simulations. First, in \S~\ref{sec:drivers}, we strive to
  understand the physical basis of variations in the SLEDs. Then, in
  \S~\ref{section:parameterization}, we use this physical
  understanding to motivate a parameterization for the SLED in terms
  of an observable quantity, the star formation surface density.  We
  show that this parameterization provides a good fit to observed
  SLEDs in \S~\ref{section:observations}.

\subsection{The Drivers of SLED Shapes in Galaxies}
\label{sec:drivers}

The principal physical drivers of CO excitation are from collisions
with \htwo, and absorption of resonant photons.  Generally, non-local
excitation by line radiation is relatively limited on galaxy-wide
scales \citep{nar08b}, and so the populations of the CO levels are set
by a competition between collisional excitation and collisional and
radiative de-excitation.  Thus the level populations in a given cloud
are driven primarily by the density and temperature in that cloud,
which affect collisional excitation and de-excitation rates, and by
the cloud column density, which determines the fraction of emitted
photons that escape, and thus the effective rate of radiative
de-excitation. In this Section we examine these factors in three
steps, first considering the gas temperature and density
distributions, then the optical depth distributions, and finally the
implications of the temperature on the shape of the SLED for
thermalized populations.

\begin{figure*}
\hspace{-1cm}
\includegraphics[scale=0.75]{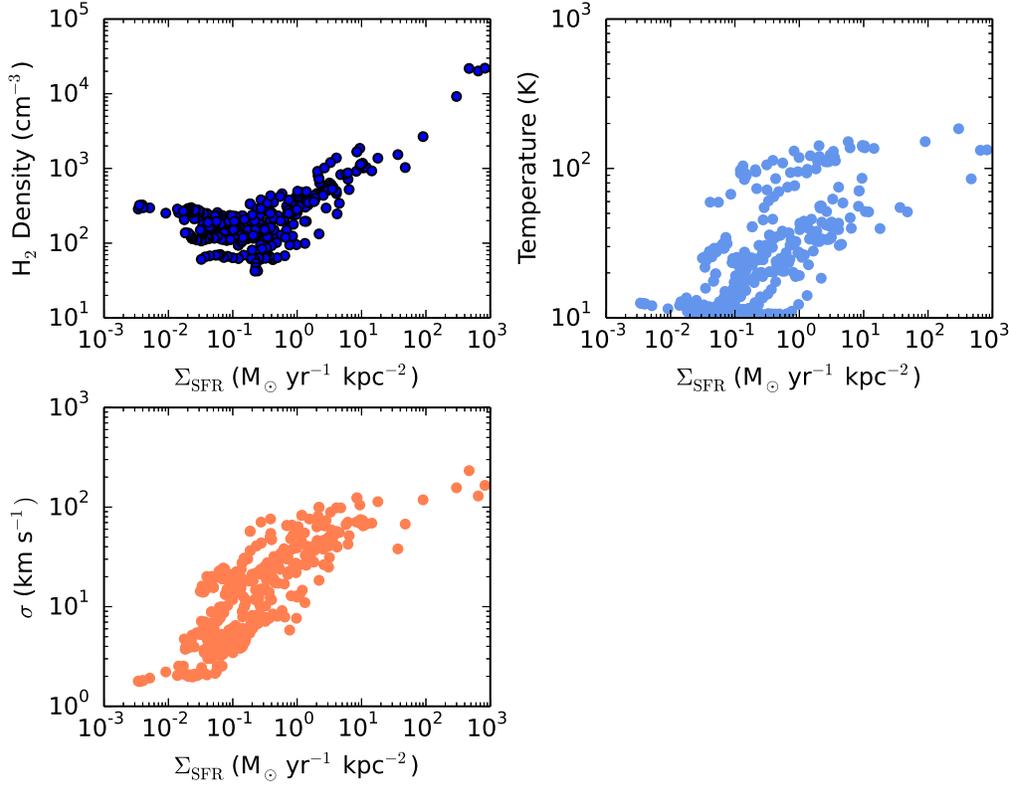}
\caption{CO mass-weighted \htwo \ gas densities, kinetic temperatures,
  and velocity dispersions of GMCs in each of our model galaxy
  snapshots as a function of galaxy $\Sigma_{\rm SFR}$. Each point
  represents a single snapshot from a single simulation. Notice that, at high
  $\Sigma_{\rm SFR}$, the physical conditions become increasingly
  extreme in our model galaxies. \label{figure:physical_quantities}}
\end{figure*}

In what follows, we will refer heavily to Figures~\ref{figure:phase_space} and
\ref{figure:physical_quantities}.
Figure~\ref{figure:phase_space} shows the distribution of GMCs in density
and temperature for two example model galaxies: a quiescent galaxy
($\Sigma_{\rm SFR} \sim 2 \times 10^{-2}$ \msun yr$^{-1}$ kpc$^{-2}$)
and a starburst ($\Sigma_{\rm SFR} \sim 650$ \msun yr$^{-1}$
kpc$^{-2}$).  For comparison, the figure also shows the temperatures corresponding to
the upper energy level of various CO transitions ($T_{\rm upper} = E_{\rm upper}/k$;
horizontal dashed lines), as well as the critical densities of the upper levels
evaluated at a gas temperature of $T=100$ K (vertical solid lines). Here we define the
critical density for a particular $J$ level as
\begin{equation}
n_{\rm crit}(J) = \frac{A_{J,J-1}}{C_{J,J-1}},
\end{equation}
where $A_{J,J-1}$ is the Einstein coefficient for spontaneous
de-excitation from level $J$ to level $J-1$, and $C_{J,J-1}$ is the
rate coefficient for collisional de-excitation between the
levels.\footnote{It is worth noting that this definition does not
  include terms associated with stimulated emission, nor does it
  include collisional de-excitation to levels other than the $J-1$
  level. Both of these are sometimes included in the denominator when
  defining the critical density.}  In Figure~\ref{figure:physical_quantities},
we plot the CO mass-weighted \htwo \ gas densities, kinetic
temperatures, and velocity dispersions of all of our model galaxies as
a function of galaxy star formation rate surface density ($\Sigma_{\rm
  SFR}$) (though note that the trends are very similar for the \htwo
\ mass-weighted quantities).  The reason for plotting $\Sigma_{\rm
  SFR}$ on the abscissa will become apparent shortly.

\subsubsection{Gas Density and Temperature}
\label{section:denstemp}

By and large, molecular clouds in normal, star-forming disk galaxies
at \zsim 0 (like the Milky Way) have relatively modest mean gas
densities, with typical values $n \approx 10-500$ \cmthree.  This is
evident from both Figures~\ref{figure:phase_space} and
\ref{figure:physical_quantities}.  Three major processes can drive
deviations from these modest values, and greatly enhance the mean gas
density of molecular clouds.  First, very gas rich systems will be
unstable, and large clumps of gas will collapse to sustain relatively
high star formation rates.  These galaxies are represented by
early-phase snapshots of both our model disks, and model mergers, and
serve to represent the modest increase in galaxy gas fraction seen in
high-\z \ star-forming galaxies \citep{nar12c, haa13,ivi13}.  Second,
galaxy mergers drive strong tidal torques that result in dense
concentrations of nuclear gas, again associated with enhanced star
formation rates \citep{mih94b,mih94a,you09,hay13c,hop13b,hop13c}.  The
mass weighted mean galaxy-wide \htwo \ gas density in the most extreme
of these systems can exceed $10^3$ \cmthree, with large numbers of
clouds exceeding $10^{4} \cmthree$ \citep[c.f. Figure A2
  of][]{nar11b}.  This is highlighted by the right panel of
Figure~\ref{figure:phase_space}, which shows the distribution of
molecular gas in the $n-$T plane for such a nuclear
starburst. Finally, large velocity dispersions in GMCs can enhance the
mean density via a turbulent compression of gas
(c.f. Equation~\ref{eq:turbulentcompression}.)  In all three cases,
the increased gas densities are correlated with enhanced galaxy-wide
SFR surface densities, in part owing to the enforcement of a
Kennicutt-Schmidt star formation law in the SPH
simulations\footnote{It is important to note that not all of the
  aforementioned drivers of enhanced gas density are causally
  correlated with increased SFR surface densities.  For example, the
  calculation of internal GMC velocity dispersions (and resultant
  turbulent compression of gas) are done in post-processing, while the
  galaxy-wide SFRs are calculated on the fly in the hydrodynamic
  simulations.  As a result, the two are not causal.}.

Increased gas kinetic temperatures can also drive up the rate of
collisions between CO and \htwo, thereby increasing molecular
excitation.  As discussed by \citet{kru11a} and \citet{nar11b}, the
gas temperature in relatively quiescent star-forming GMCs is roughly
$\sim 10$ K.  In our model, this principally owes to heating by cosmic
rays.  At the relatively low densities ($10-100 \ \cmthree$) of normal
GMCs, heating by energy exchange with dust is a relatively modest
effect, and cosmic rays tend to dominate the gas heating.  For a
cosmic ray ionization rate comparable to the Milky Way's, a typical
GMC temperature of $\sim 10$ K results \citep{nar13}\footnote{This is what
drives the apparent low temperature cut-off in the left panel of
Figure~\ref{figure:phase_space}, and top-right of
Figure~\ref{figure:physical_quantities}.}.  

As galaxy star formation rates increase, however, so do the gas
temperatures.  This happens via two dominant channels.  In our
models, the cosmic ray ionization rate scales linearly with the
galaxy-wide star formation rate, motivated by tentative trends
observed by \citet{acc09}, \citet{abd10b}, and \citet{hai08}.  As noted by
\citet{pap10a}, \citet{pap10b}, and \citet{nar12b,nar13b}, increased cosmic ray
heating rates can substantially drive up gas kinetic temperatures,
depending on the density of the gas and cosmic ray flux.  Second, at
gas densities above $\sim 10^{4} \ \cmthree$, the energy exchange
between gas and dust becomes efficient enough that collisions between
\htwo \ and dust can drive up the gas temperature as well.  Because
high star formation rates accompany high gas densities, and high dust
temperatures typically accompany high star formation rates
\citep[e.g.][]{nar10b}, the gas temperature rises with the
galaxy SFR.  At very large densities ($n \gg 10^{4} \cmthree$) we find
that the dust temperatures are typically warm enough that the heating
by cosmic rays is a minor addition to the energy imparted to the gas by
collisions with dust.  As is evident from
Figure~\ref{figure:physical_quantities}, because both heating channels
scale with the observable proxy of SFR surface density, the mean
galaxy-wide gas temperature does as well.  For galaxies undergoing
intense star formation, the mean gas temperature across the galaxy can
approach $\sim 100 $ K.

\subsubsection{Line Optical Depths}
\label{section:optical_depths}

\begin{figure}
\includegraphics[scale=0.45]{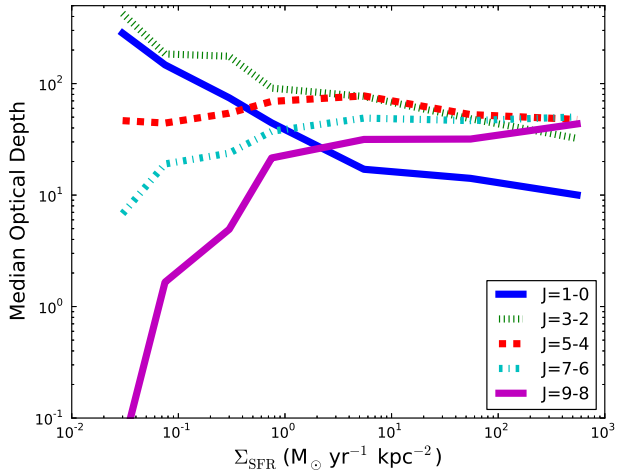}
\caption{CO line luminosity-weighted line optical depths as a function
  of galaxy $\Sigma_{\rm SFR}$.  Each line represents the median
  luminosity-weighted line optical depth for all snapshots within bins
  of $\Sigma_{\rm SFR}$.  The low-$J$ CO lines start off roughly
  thermalized.  At increased $\Sigma_{\rm SFR}$, the gas velocity
  dispersions increase, and photons in these lines are more easily able to escape,
  decreasing the optical depths.  For high-$J$ transitions, the level
  populations begin at modest levels, and the optical depths are low.
  Broadly, at increased $\Sigma_{\rm SFR}$, the level populations rise
  owing to increased gas densities and temperatures, which drives the
  optical depths up.\label{figure:tauplot}}
\end{figure}

From \S~\ref{section:denstemp}, it is clear that at increased star
formation rate surface densities, the gas temperatures and densities
rise, which will drive up molecular excitation.  The next question is,
how quickly will these levels approach LTE, and saturate the observed
SLED?  The critical density provides a rough guide to this, but
in optically thick regions, line trapping can increase excitations at
a level above what would be expected if the photons could escape
easily \citep{eva99,rei11}.  In the escape probability approximation, we
can define an effective critical density
\begin{equation}
n_{\rm crit,eff}(J) = \frac{\beta_{J,J-1} A_{J,J-1}}{C_{J,J-1}}
\end{equation}
where $\beta_{J,J-1}$ is the probability that a photon emitted at a
frequency corresponding to the energy difference between levels
$J$ and $J-1$ will escape the cloud without being absorbed.
Recall from Equation~\ref{eq:kmt_beta} that $\beta_{J,J-1}$ can be
approximated by $1/[1+(3/8) \tau_{J,J-1}]$ \citep{dra11}, so
for $\tau_{J,J-1}\gg 1$, the effective density needed to thermalize
a given level decreases by roughly a factor of $\tau_{J,J-1}$.

In Figure~\ref{figure:tauplot}, we show the median CO
luminosity-weighted optical depth for a sample of CO transitions as a
function of galaxy star formation rate surface density.  The lines in
Figure~\ref{figure:tauplot} can be thought of as the typical optical
depth for different CO lines within the GMCs in our model galaxies.
In short, low-$J$ CO transitions are always optically thick, while
high-$J$ lines only become optically thick in more extreme star-forming
galaxies.  Referring to Equation~\ref{eq:kmt_tau}, the qualitative
trends are straightforward to interpret, and arise from a combination
of both the varying level populations, as well as changing physical
conditions as a function of galaxy $\Sigma_{\rm SFR}$.  Low lying
transitions such as CO $J=1- 0$ are typically thermalized in most
clouds in galaxies (even at low $\Sigma_{\rm SFR}$), owing to the
relatively low critical densities and high optical depths.  Hence, in
even more dense conditions in high $\Sigma_{\rm SFR}$ systems, the
relative level populations stay approximately constant. However, the
optical depth is inversely proportional to the gas velocity dispersion
which tends to rise with star formation rate surface density
(c.f. Figure~\ref{figure:physical_quantities}).  Hence, at increased
$\Sigma_{\rm SFR}$, the optical depths of low-$J$ transitions decreases
slightly, though they remain optically thick.  The opposite trend is true
for high-$J$ transitions.  These transitions are typically subthermally
populated in low density clouds.  As densities and temperatures rise
with $\Sigma_{\rm SFR}$, so do the level populations.  This rise
offsets the increase in gas velocity dispersion, and the optical
depths in these lines increase in more extreme star forming systems.

The net effect of the mean trends in CO optical depths in molecular
clouds in star forming galaxies, as summarized in
Figure~\ref{figure:tauplot}, is that galaxies of increased gas
density and temperature (as parameterized here by increased
$\Sigma_{\rm SFR}$) have increased optical depths in high-$J$ CO lines.
This drives down the effective density, and lets the levels achieve
LTE more easily.  This will be important for explaining how some
observed high $J$ CO lines (such as $J=7-6$) can exhibit line ratios
comparable to what is expected from LTE, despite the fact that 
galaxy-wide mean densities rarely reach $\sim 10^5$ $\cmthree$, the typical
critical density for high-$J$ CO emission
(c.f. Figure~\ref{figure:phase_space}).
 
\subsubsection{Line Ratios for Thermalized Populations}
\label{section:thermalized}

 Thus far, we have been principally concerned with the physical drivers
 of CO excitation, as well as the influence of optical depth in
 lowering the effective density for thermalization of a given CO line.
 These have given some insight as to roughly when SLED lines will
 saturate.  We now turn our attention to what values line ratios will
 have when they approach thermalization.  The intensity of a line for
 level populations that are in LTE is given by the Planck function at that
 frequency.  In the Rayleigh-Jeans limit, when $E_{\rm upper} \ll kT$, the
 observed line ratios (compared to the ground state) scale as $J^2$.  

\begin{figure}
\includegraphics[scale=0.45]{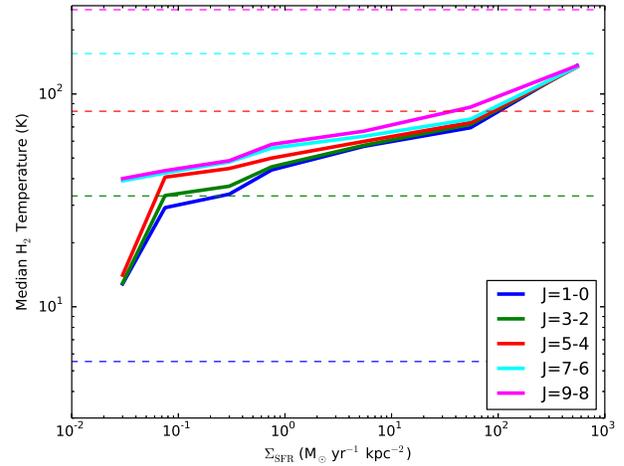}
\caption{CO line luminosity-weighted median gas kinetic temperatures as a
  function of galaxy $\Sigma_{\rm SFR}$.  The horizontal dashed lines
  show the temperature $T_{\rm upper} = E_{\rm upper}/k$
  that corresponds to the upper level energy of a
  given CO transition (with the same color coding as the solid
  lines).  At even modest $\Sigma_{\rm SFR}$, gas temperatures as such that lines
  through CO $J=4-3$ are well into the Rayleigh-Jeans limit. In contrast,
  even at the most extreme $\Sigma_{\rm SFR}$ values in our library,
  the median gas temperature is never high enough to exceed
  $T_{\rm upper}$ for $J_{\rm upper} \geq 7$.
  \label{figure:line_weighted_temp}}
\end{figure}

In Figure~\ref{figure:line_weighted_temp}, we plot the CO line
luminosity-weighted gas kinetic temperature as a function of galaxy
star formation rate surface density for a selection of CO transitions.
To guide the eye, we also plot as horizontal (dashed) lines the values
of $T_{\rm upper} = E_{\rm upper}/k$ for the same transitions.  At even low values
of $\Sigma_{\rm SFR}$, the CO $J=1- 0$ and $3-2$ emitting gas temperature
is a factor $\sim 5-10$ larger than $E_{\rm upper}/k$.  The gas that
emits higher-lying lines (e.g. $J=5-4$) has temperatures roughly twice
$E_{\rm upper}/k$ for the most extreme systems, while higher lying
lines do not approach the Rayleigh-Jeans limit even for the most
extreme galaxies.  As a consequence, for high $\Sigma_{\rm SFR}$
galaxies, we can expect line ratios (compared to the ground
state) to scale as $J^2$ for lower-lying lines.  For higher $J$
lines in LTE ($J_{\rm upper} \ga 7$), the line ratios will simply
follow ratio of the Planck functions for the levels, giving rise to
a SLED that is flatter than $J^2$.

\subsection{A Parameterized Model for CO SLEDs}
\label{section:parameterization}

Owing to both a desire for high-resolution and sensitivity, as well as
the general prevalence of (sub)mm-wave detectors that operate between
$800$ $\micron$ and $\sim 3$ mm, observations of high-\z \ galaxies are
often of rest-frame high-$J$ CO lines.  Because physical measurements of
\htwo \ gas masses require conversion from the CO  $J=1- 0$ line, we are
motivated to utilize our model results to derive a parameterized model
of the CO excitation in star forming galaxies.

From Figure~\ref{figure:cosled_physical}, it is clear that the SLED
from galaxies rises with increasing \htwo \ gas density and kinetic
temperature.  How rapidly these SLEDs rise depend on the exact shape
of the density and temperature probability distribution function
(PDF), as well as the line optical depths.  As we showed in
\S~\ref{section:thermalized}, the SLED tends to saturate at roughly
$J^2$ for $J_{\rm upper} \la 6$ at high $\Sigma_{\rm SFR}$, owing to
large gas kinetic temperatures
(c.f. Figure~\ref{figure:line_weighted_temp}).  For higher lying
lines, when the levels are in LTE, the SLED will saturate at the ratio
of the Planck functions between $J_{\rm upper}$ and the ground state. 

In principle, the SLED would best be parameterized by a combination of
the gas densities, temperatures and optical depths.  Of course, these
quantities are typically {\it derived} from observed SLEDs, not
directly observed.  However, as shown in
Figures~\ref{figure:phase_space}-\ref{figure:tauplot}, these
quantities are reasonably well-correlated with the star formation rate
surface density of galaxies.  Qualitatively this makes sense - more
compact star forming regions arise from denser gas concentrations.
Similarly, regions of dense gas and high star formation rate density
have large UV radiation fields (and thus warmer dust temperatures),
increased efficiency for thermal coupling between gas and dust, and
increased cosmic ray heating rates.

\begin{table}
\caption{Fits to Equation~\ref{eq:fit} for Resolved Observations of
  Galaxies.}
\label{table:resolved_fit}
\begin{center}
\begin{tabular}{|l|l|l|l|}
\hline\hline
{\bf Line Ratio} & {\bf A} & {\bf B} &  {\bf C} \\
($I_{\rm ij}/I_{\rm 1-0}$) &&\\
\hline
\hline
CO $J=2-1$ & 0.09 &0.49 &0.47 \\
CO $J=3-2$ & 0.17 &0.45 &0.66 \\
CO $J=4-3$ & 0.32 &0.43 &0.60 \\
CO $J=5-4$ & 0.75 &0.39 &0.006 \\
CO $J=6-5$ & 1.46 &0.37 &-1.02 \\
CO $J=7-6$ & 2.04 &0.38 &-2.00 \\
CO $J=8-7$ & 2.63 &0.40 &-3.12 \\
CO $J=9-8$ & 3.10 &0.42 &-4.22 \\
\hline
\hline
\end{tabular}
\end{center}
\end{table}

\begin{table}
\caption{Fits to Equation~\ref{eq:fit} for Unresolved Observations of
  Galaxies.}
\label{table:unresolved_fit}
\begin{center}
\begin{tabular}{|l|l|l|l|}
\hline\hline
{\bf Line Ratio} & {\bf A} & {\bf B} &  {\bf C} \\
($I_{\rm ij}/I_{\rm 1-0}$) &&\\
\hline
\hline
CO $J=2-1$ & 0.07 & 0.51 & 0.47 \\
CO $J=3-2$ & 0.14 & 0.47 & 0.64 \\
CO $J=4-3$ & 0.27 & 0.45 & 0.57 \\
CO $J=5-4$ & 0.68 & 0.41 & -0.05 \\
CO $J=6-5$ & 1.36 & 0.38 & -1.11 \\
CO $J=7-6$ & 1.96 & 0.39 & -2.14 \\
CO $J=8-7$ & 2.57 & 0.41 & -3.35 \\
CO $J=9-8$ & 2.36 & 0.50 & -3.70 \\

\hline
\hline
\end{tabular}
\end{center}

\end{table}

\begin{figure}
\includegraphics[scale=0.45]{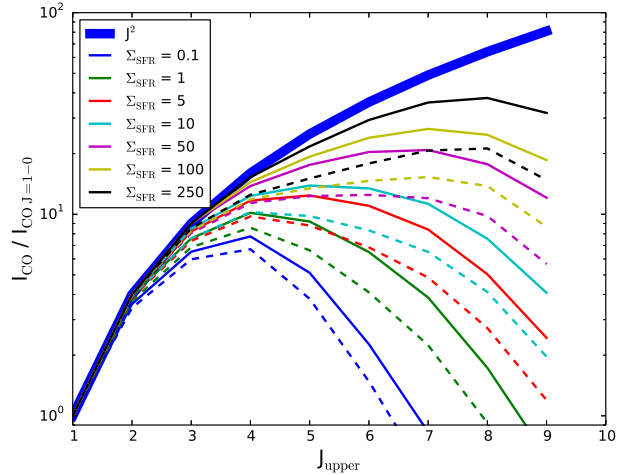}
\caption{Parameterized model for SLEDs. The model SLED shape is a
  function of $\Sigma_{\rm SFR}$, and is given in
  Equation~\ref{eq:fit}, with fit parameters listed in
  Tables~\ref{table:resolved_fit} and \ref{table:unresolved_fit}. 
  The solid lines show the perfectly resolved case, while the dashed
  lines (of the same color) show the case for unresolved models. 
  Unresolved observations share the same assumed emitting area, and
  therefore depress the SLEDs at high-$J$ because the emitting gas occupies
  a smaller area, and therefore suffers beam dilution when averaged
  over the same area as the low-$J$ lines.  \label{figure:cosled_sigma_sfr}}
\end{figure}

Motivated by these trends, we fit the simulated line ratios for all
model snapshots in our galaxies as a power law function of the star
formation rate surface density utilizing a Levenberg-Marquardt
algorithm. The power-law fit encapsulates the behaviour expected for
line intensity ratios (with respect to the ground state) as a function
of $\Sigma_{\rm SFR}$: at low $\Sigma_{\rm SFR}$, the line ratios rise
as the levels approach LTE, then plateau as the levels are
thermalized at high $\Sigma_{\rm SFR}$.  Our model line ratios are
well fit by a function:
\begin{equation}
\label{eq:fit}
{\rm log_{10}}\left[ \frac{I_{\rm ij}}{I_{\rm 1-0}}\right] = A \times \left[{\rm log_{10}}(\Sigma_{\rm SFR})-\chi\right]^B + C
\end{equation}
where the $\Sigma_{\rm SFR}$ is in \msun yr$^{-1}$ kpc$^{-2}$, and the
offset $\chi = -1.85$; this value is chosen to ensure that our fit
produces purely real values for $I_{\rm ij}/I_{1-0}$, and is
determined primarily by the range of $\Sigma_{\rm SFR}$ values sampled
by our simulation library. We list the best fit values for model
parameters $A$, $B$, and $C$ in Table~\ref{table:resolved_fit}, and
show how the functional form of the best-fit SLED varies with
$\Sigma_{\rm SFR}$ with the solid lines in
Figure~\ref{figure:cosled_sigma_sfr}.  The parameterized form given by
Equation~\ref{eq:fit} is valid for SFR surface densities $\Sigma_{\rm
  SFR} \geq 1.5 \times 10^{-2}$ \msunyr \ kpc$^{-2}$, as this is the
range above which our fits are well-behaved.  Because our fits derive
from luminosity-weighted quantities, the fits are valid down to our
resolution limit.  It is worth noting that the line ratios from the
highest $J$ lines do not approach $J^2$, even for extremely high
$\Sigma_{\rm SFR}$ galaxies.  This represents a combination of both
gas not being in the Rayleigh-Jeans limit
(c.f. \S~\ref{section:thermalized}), as well as (at times) not in LTE.

In principle, the numbers presented in Table~\ref{table:resolved_fit}
allow for the calculation of the full SLED of a galaxy given a
measured $\Sigma_{\rm SFR}$. This said, an essential point to the SLED
calculations presented thus far is that CO intensities for a given line
are resolved at the level of our finest cell in our adaptive mesh ($\sim
70$ pc).  Thus, Table~\ref{table:resolved_fit} provides reliable
parameters for a SLED construction from the SFR surface density given
that the emitting region in each line is resolved.  In other words, if
a given high-$J$ line is observed from a galaxy (say, CO $J=4-3$), to
convert down to the $J=1-0$ transition utilizing Equation~\ref{eq:fit} and
the fit parameters in Table~\ref{table:resolved_fit}, the CO $J=4-3$
region must have been resolved.  For observations of local galaxies
with interferometers (e.g. ALMA), this is plausible, and even likely.
For high-redshift observations, or observations of local
galaxies using coarse beams, however, it is likely that the
observations will be unresolved.

In the case of unresolved observations, the emitting area utilized in
intensity calculations is typically a common one across transitions,
and therefore misses the clumping of the principle emitting regions of
higher-lying CO transitions.  In this case,
Table~\ref{table:resolved_fit} will not provide an appropriate
description of the observed SLED.  To account for this, we determine
the luminosity-weighted emitting area for each transition of each
model galaxy snapshot, and scale the intensity of the line
accordingly.  We then re-fit the line ratios as a function of
$\Sigma_{\rm SFR}$, and present the best fit parameters for unresolved
observations in Table~\ref{table:unresolved_fit}.  The effect of an
unresolved observation is clear from
Figure~\ref{figure:cosled_sigma_sfr}, where we show the unresolved
counterparts to the resolved SLEDs with the dashed curves.  For a
given transition, the solid line shows the resolved SLED from our
model parameterization, and the dashed the unresolved excitation
ladder.  Because high $J$ lines tend to arise from warm, dense regions
that are highly clumped compared to the CO $J=1- 0$ emitting area, the
resultant SLED is decreased compared to highly resolved observations.
This effect was previously noted for models high-\z \ Submillimeter
Galaxy models by \citet{nar09}.

In conjunction with this paper, we provide IDL and Python modules for
reading in the parameterized fits from Tables~\ref{table:resolved_fit}
and \ref{table:unresolved_fit}, and host these at a
\href{http://desika.haverford.edu/codes/}{\texttt{dedicated web
    page}}, as well as at \url{https://sites.google.com/a/ucsc.edu/krumholz/codes}.

\subsection{Comparison to Observations}
\label{section:observations}

\begin{figure*}
\begin{centering}
\includegraphics[scale=0.95]{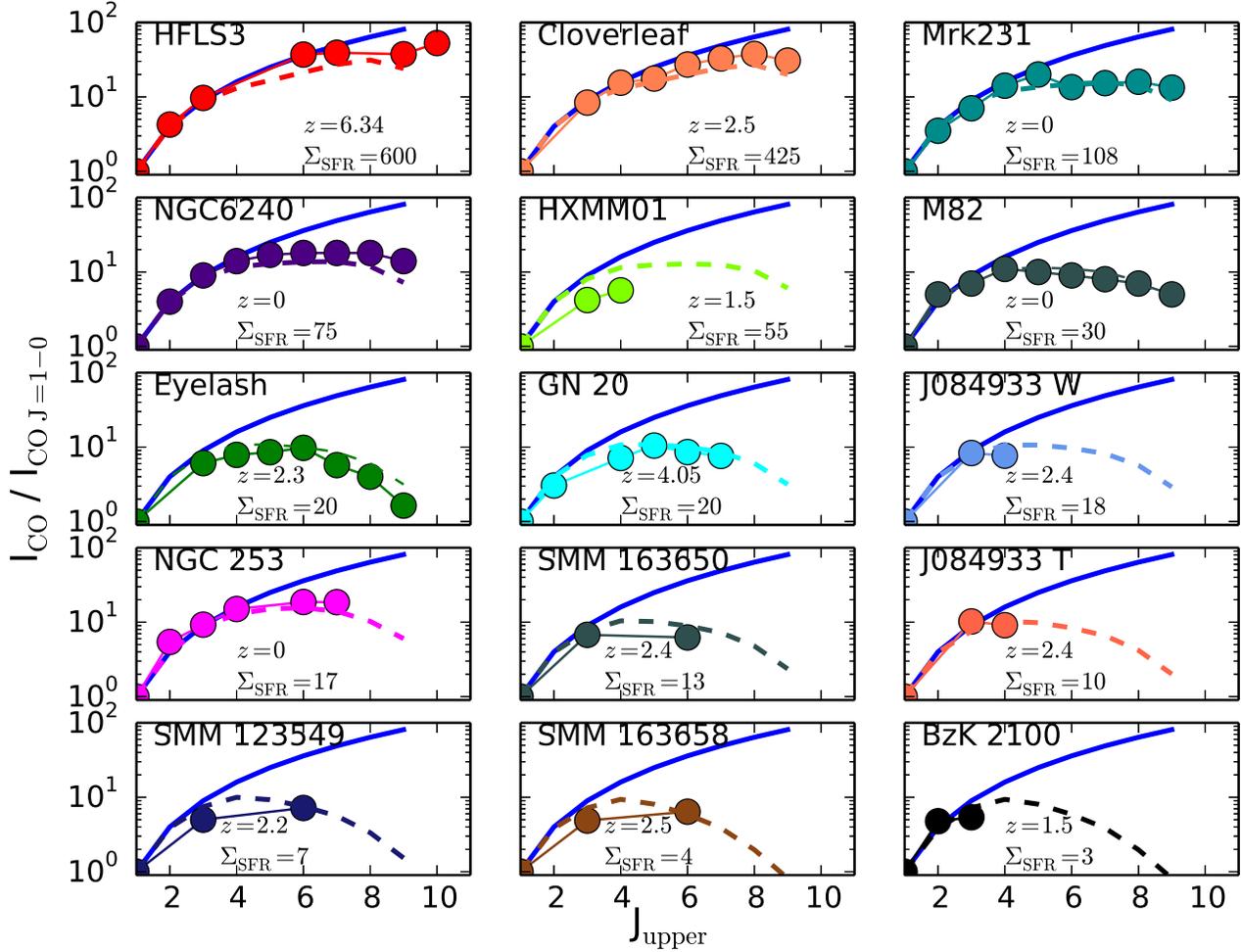}
\end{centering}
\caption{Compilation of fifteen observed CO Spectral Line Energy
  Distributions (solid colored points) for galaxies at low and
  high-\z.  The panels are sorted in order of decreasing $\Sigma_{\rm
    SFR}$, and span a dynamic range of roughly 200. Each panel is
  labeled by the value of $\Sigma_{\rm SFR}$, in units of $\msun$
  kpc$^{-2}$ yr$^{-1}$, and the redshift of the galaxy.  The dashed
  line in each panel shows the model prediction derived in this paper
  (Equation~\ref{eq:fit} and Table~\ref{table:unresolved_fit} for all
  galaxies but NGC 253, Equation~\ref{eq:fit} and Table
  \ref{table:resolved_fit} for NGC 253, which is spatially resolved).
  The solid blue lines show $I \propto J^2$, the scaling expected for
  thermalized populations in the Rayleigh-Jeans limit.  The data shown
  are from the following sources:
  \citet{rie13,rie11a,bar94,wil95b,bar97,wei03,all97,yun97,bra09,fu13,dan11,cas09b,car10,dad09,dan09,hod12,kam12,ivi13,bra03,hai08b,ivi11,dad10a,mei13,pap12b,van10b}.
  The bulk of the $\Sigma_{\rm SFR}$ values were either reported in
  the original CO paper, or in \citet{ken98b}. The exceptions are as
  follows.  The value of $\Sigma_{\rm SFR}$ for the Eyelash galaxy and
  GN 20 were provided via private communication from Mark Swinbank and
  Jackie Hodge, respectively.  For SMM 123549, 163650 and 163658,
  $\Sigma_{\rm SFR}$ values were calculated by converting from L$_{\rm
    FIR}$ to SFR utilizing the calibration in \citet{mur11} and
  \citet{ken12}, and the galaxy sizes reported in \citet{ivi11}.  The
  size scale for the Cloverleaf quasar was taken from the CO $J=1- 0$
  measurements by \citet{ven03} for the $\Sigma_{\rm SFR}$
  calculation, and SFR reported in \citet{sol03}. The $\Sigma_{\rm
    SFR}$ value for Mrk231 was calculated by via the SFR reported for
  the central disk by \citet{car98}, and the size of the disk derived
  by \citet{bry96}.
\label{figure:sled_multiplot}
  }
\end{figure*}

In an effort to validate the parameterized model presented here, in
Figure~\ref{figure:sled_multiplot} we present a comparison between
our model for CO excitation in galaxies and observations of 15
galaxies from present epoch to $z>6$.  Most of the observational data
comes from galaxies at moderate to high-redshift owing to the fact
that it is typically easier to access the high-$J$ submillimeter CO
lines when they are redshifted to more favorable atmospheric
transmission windows than it is at \zsim 0.  With the exception of NGC
253, we use our fits for unresolved observations in this comparison.
For NGC 253 we use the resolved model, because that galaxy is only
$\approx 3.3$ Mpc away \citep{rek05}, and the CO observations were
made with fairly high resolution, so that the high-$J$ emitting regions
were likely resolved. The blue solid lines in each panel show the $J^2$ scaling
of intensities expected for thermalized levels in the Rayleigh-Jeans limit.

Figure \ref{figure:sled_multiplot} shows that the model performs quite
well over a very large dynamic range in galaxy properties.  Over all
transitions in all galaxies in Figure~\ref{figure:sled_multiplot}, the
median discrepancy between observations and our model is $25\%$, and
the maximum ratio is a factor $\sim 2$.  In comparison, for the
observed set of SMGs presented in
Figure~\ref{figure:cosled_observational}, the maximum difference in
line ratios at a given transition is $\sim 30$.  The galaxies shown
range from local starbursts to relatively normal star-forming galaxies
at high-\z \ to high-\z \ SMGs to some of the most luminous,
heavily-star forming galaxies discovered to date (e.g. GN20 and
HFLS3).  Their star formation surface densities vary by roughly a
factor of 200. Our model captures both the qualitative and
quantitative behavior of the SLEDs over this full range.

\begin{figure}
\includegraphics[scale=0.45]{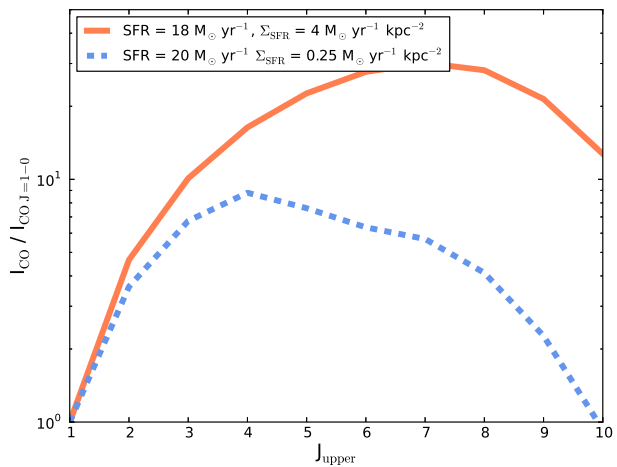}
\caption{Simulated resolved CO SLEDs for two model galaxies with similar star
  formation rates, but differing star formation rate surface densities
  ($\Sigma_{\rm SFR}$).  While two galaxies may have the same global
  SFR, the physical conditions in the ISM may vary strongly leading to very
  different SLEDs. The star formation surface density $\Sigma_{\rm
    SFR}$ provides a much better parameterization. \label{figure:sfr_sigma_comparison}}
\end{figure}

While the star formation rate surface density provides a relatively
good parameterization for the observed SLED from a star forming
galaxy, the global SFR does not.  This is because the global SFR does
not correlate well with the density and temperature PDFs in clouds.
As an example, in Figure~\ref{figure:sfr_sigma_comparison} we plot the
(resolved) simulated SLED for two galaxies with very similar global
SFRs.  One is a galaxy merger, and the other a gas-rich disk galaxy
comparable to conditions in high-\z \ disks.  The galaxy merger is
more compact, and therefore has a significantly higher $\Sigma_{\rm
  SFR}$ than the relatively distributed star formation characteristic
of the gas-rich disk.  As a result, though the global star formation
rates of the galaxies are similar, the physical conditions within the
molecular clouds of the systems are not, and the resultant observed
SLEDs diverge. Similar divergences can be found in observations.  For
example, compare the Cloverleaf quasar to the Submillimeter Galaxy SMM
163650 in Figure~\ref{figure:cosled_observational}.  Both galaxies
have incredible global SFRs of $\sim 1000 \ \msunyr$
\citep{sol03,wei03,ivi11}, but they clearly have starkly different CO
SLEDs\footnote{Though note that if differential magnification for the
  Cloverleaf is important, this may complicate the comparison.}.

\section{Discussion}
\label{section:discussion}

\subsection{The Parameterization of SLEDs}

We have presented a model the physical origins of SLEDs in star forming galaxies, and
derived a rough parameterization of the SLED in terms of an observable quantity,
the star formation rate surface density.  As
discussed in \S~\ref{section:parameterization}, as well as in
Figure~\ref{figure:sfr_sigma_comparison}, the global SFR alone
provides a poor parameterization of the CO SLED, and more generally,
the physical conditions in the molecular clouds in the galaxy.

A number of other works have arrived at similar conclusions, though
via notably different methods.  As an example, \citet{ruj11} showed
that galaxies with similar infrared luminosity densities across
different redshifts shared more common observed properties (such as
aromatic features) than simply galaxies at a shared infrared
luminosity.  This is akin to the idea that ULIRGs at \zsim 0 do not
lie on the SFR-$M_*$ main sequence, while galaxies of a comparable
luminosity at high-\z \ do \citep{dad05,rod11}.  In contrast,
normalizing by the spatial scale of the emitting (or star-forming)
region allows for a more straightforward comparison of galaxies even
at vastly different redshifts. A similar point was made by
\citet{hop10}, who showed that with increasing redshift, the fraction
of galaxies at a given luminosity comprised of galaxy mergers and
starbursts decreases dramatically.

This point is related to that made by \citet{dia13}, when examining the
[C~\textsc{ii}]-FIR deficit in galaxies.  At \zsim 0, galaxies with large
infrared luminosity tend to have depressed [C~\textsc{ii}] 158 \micron
\ emission as compared to what would be expected for the typical
           [C~\textsc{ii}]-FIR scaling at lower luminosities
           \citep{mal97b,luh98,del11b,far13}.  While the origin of the
           deficit is unknown, early indications suggest that a
           similar deficit may exist in high-\z \ galaxies, though
           offset in infrared space \citep[e.g.][]{sta10,gra11,wan13}.
           However, \citet{dia13} showed that when comparing to a
           galaxy's location with respect to the specific star
           formation rate main sequence, galaxies at both low and
           high-\z \ follow a similar trend.

As a result of this, there are no ``typical" values of CO SLEDs for a
given galaxy population.  Populations such as ULIRGs or SMGs are
generally selected via a luminosity or color cut.  This
can result in a heterogeneous set of physical conditions in galaxies,
and therefore heterogeneous CO emission properties.  This was shown
explicitly by \citet{pap12b}, who demonstrated that local luminous infrared
galaxies (LIRGs; \lir$\geq 10^{11}$ \lsun) exhibit a diverse range of
CO excitation conditions.  Similarly, \citet{dow98},
\citet{tac08}, \citet{mag11}, \citet{hod12}, \citet{mag12}, \citet{ivi13},
and \citet{fu13} showed that the CO-\htwo
\ conversion factor, \xco, for ULIRGs and high-\z \ SMGs each show a
factor $\sim 5-10$ range in observationally derived values.  A
template for the CO emission properties is not feasible for a given
galaxy population given the wide variation in ISM physical conditions
that can be found in luminosity-selected populations.

\subsection{Relationship to Other Models}

Over the last few years, a number of theoretical groups have devoted
significant attention to predicting CO emission from star-forming
galaxies.  Here, we review some previous efforts to predict CO SLEDs,
and place our current work into this context.  In
short, a number of analytic and semi-analytic models have explored the
simulated SLEDs from model galaxies, but no comprehensive model for
the origin of SLEDs has been reported thus far.  Most works focus on
validation of a given galaxy formation model.

To date, no work investigating SLEDs has utilized bona fide
hydrodynamic simulations of galaxy evolution that directly simulate
the physical properties of the interstellar medium to calculate the CO
excitation from galaxies. They have instead relied on either analytic
or semi-analytic models to describe the physical properties of
galaxies and giant molecular clouds within these galaxies, and coupled
these analytic models with numerical radiative transfer calculations
to derive the synthetic CO emission properties.

On the purely analytic side, \citet{arm12} and \citet{mun13} developed
models for quiescent disk galaxy evolution (constructing such a model
for merger-driven starbursts would be quite difficult)
in order to predict CO emission from nuclear starburst disks around AGN,
and from \z$\ga 6$
galaxies, respectively.  \citet{arm12} construct single phase nuclear
disks that rotate around a central AGN.  The SFR is forced to be such
that Toomre's $Q$ parameter is unity, and the vertical support of the
disk owes to radiative feedback following the theory of \citet{tho05}.
The CO SLEDs are calculated by applying {\sc ratran} radiative
transfer calculations to the model grid \citep{hog00}.  These authors
find that nuclear starbursts are extremely dense, and therefore
thermalized line ratios will be common.  A subset of the models show
strong correspondence with the M82, Mrk 231 and Arp220, with
potentially some under-prediction of the highest $J$ transitions ($J>8$).
Interestingly, a subset of the models by \citet{arm12} fall into CO
line absorption at large $J_{\rm upper}$ ($J_{\rm upper} > 11$), owing
to the steep temperature and density gradients within the starburst.

\citet{mun13} develop analytic models of high-\z \ galaxy disks that
also are forced to maintain vertical hydrostatic equilibrium by
radiation pressure, similar to the aforementioned study by
\citet{arm12}.  \citet{mun13} place their galaxies in a cosmological
context by allowing for gas accretion from the intergalactic medium
(IGM) following parameterizations from cosmological dark matter
simulations, as well as allow for momentum driven winds following the
\citet{opp08} prescriptions.  Within the disks, GMCs are assumed to
have a threshold floor surface density of 85 \msun pc$^{-2}$, and
\htwo \ fractions that depend on the stellar and cloud surface
density.  The CO fraction follows the \citet{wol10} model, as in our
work, and the GMCs have lognormal density distribution following from
supersonic turbulence. After accounting for the modulation of critical
densities by line optical depths, \citet{mun13} calculate the CO
excitation of their model \zsim 6 galaxies utilizing a precursor code
to \despotic, with identical underlying algorithms.  These authors
focused specifically on the observable CO $J=6-5$ line.  They find
that the high $J$ lines from these galaxies will be observable with
ALMA, as long as the ISM clumps similarly to local galaxies.  If
turbulence does not provide the principal support for the clouds,
however, and they are more homogenous, and the excitation of high $J$
levels will be greatly reduced.  This said, \citet{mun13b} suggest
that their model may overpredict the expected CO $J=6-5$ emission from
extreme star-formers at high-\z \ such as the \zsim 6 SMG HFLS3, in
part due to the high gas velocity dispersions (and thus an increased
gas median density due to turbulent compression of gas;
c.f. equation~\ref{eq:turbulentcompression}).  Indeed their model has
similar velocity dispersions for HFLS3-like systems as our own ($\sim
100$ $\kmsend$).  While it is difficult to exactly compare our model
with that of \citet{mun13b} owing to the starkly different modeling
methods (analytic versus coupled hydrodynamic and radiative transfer),
one possible origin in the difference in predicted SLEDs for
HFLS3-like systems is in the radial density distribution within the
gas.  In our model, a sufficiently large low-density emitting-region
exists even in high $\Sigma_{\rm SFR}$ galaxies that the unresolved
model SLED is depressed at high $J$ levels.

Turning to semi-analytic models, \citet{lag12} coupled the {\sc
  galform} SAM with the 1D photodissociation region (PDR) code {\sc
  ucl\_pdr} \citep{bay11} to analyze the model CO emission from
analytic galaxies painted on to dark matter halos taken from N-body
simulations.  These authors assume \htwo \ fractions that derive from
the empirical relations provided by \citet{bli06} for normal
star-forming galaxies\footnote{While the \citet{bli06} empirical
  relation provides a reasonable match to observational constraints at
  solar metallicity, \citet{fum10} and \citet{won13} show that the
  relation may provide a more poor estimate of \htwo \ gas fractions
  at low metallicity.}, and an \htwo \ fraction of unity in
starbursts.  The PDR modeling assumes the gas is confined to a 1-zone
slab geometry illuminated on one side by a UV field, which scales with
$\Sigma_{\rm SFR}$ in the \cite{lag11,lag12} version of \galform.  For
the majority of their models, \citet{lag12} assume isodensity gas of
$n_{\rm H} = 10^{4} \cmthree$, and calculate the SLEDs accordingly.
They find that on average the SLEDs of galaxies that are LIRGs(ULIRGs)
at $z=0$ peak at $J=4(5)$, though there is significant dispersion in
the high-$J$ excitation.  While difficult to directly compare with our
own work, the range of SLEDs derived by \citet{lag12} is certainly
consistent with the range found here.

Following on this, \citet{pop13b} and \citet{pop13} coupled a
semi-analytic model for galaxy formation \citep{som08} with the
radiative transfer code $\beta${\sc3D} \citep{poe05,poe06,per11} to
model the CO (among other species) emission from their model galaxies.
The \htwo \ fraction is based on fitting functions from cosmological
zoom simulations by \citet{gne11}, dependent on the stellar and gas
surface densities.  Star formation follows a modified
Kennicutt-Schmidt star formation relation \citep{ken98b,ken12}, and
clouds occupy a volume filling factor of $f_{\rm clump} = 0.01$.
These authors find excellent agreement in the dense gas fractions
returned by their models \citep[as traced by observed dense gas line
  ratios --][]{nar05,bus08,jun09,hop13} and observations.  Relevant to
the present study, \citet{pop13b} find that galaxies with higher FIR
luminosities have CO SLEDs that peak at higher excitation levels,
similar to the work by \citet{lag12}.  Galaxies at a fixed FIR
luminosity in this model, though, peak at higher CO levels at high
redshifts than at low redshifts.  In the \citet{pop13b} model, this
owes to the fact that galaxies at high-\z \ at a fixed FIR have warmer
and denser cold gas than their low-\z \ counterparts do.  While we are
unable to make a similar comparison, in our model this could be
ascribed to galaxies at high-\z \ at a fixed luminosity having higher
$\Sigma_{\rm SFR}$ than their low-\z \ analogs.  This is likely the
case in the \citet{pop13b} model, as the disk sizes of galaxies at a
fixed stellar mass are smaller at high-\z \ than present epoch
\citep{som08b}.  \citet{pop13b} show a steeply rising relation between
CO level populations and $\Sigma_{\rm gas}$ (which is presumably
similar to the relation with $\Sigma_{\rm SFR}$ given their star
formation law), and a much more moderate relationship between
molecular level populations and global galaxy SFR.  In this sense our
work is seemingly in very good agreement with the \citet{pop13b}
model.

\subsection{Applications of this Model}
With the powerlaw fit presented in Equation~\ref{eq:fit} and
Tables~\ref{table:resolved_fit} and ~\ref{table:unresolved_fit}, we
have provided a mechanism for calculating the full CO rotational
ladder through the $J=9-8$ transition given a constraint on the star
formation rate surface density.  With this, in principle one can
utilize a CO observation of a high-\z \ galaxy at an arbitrary
transition, and then down-convert to the CO $J=1- 0$ transition flux
density.

To convert from the CO $J=1- 0$ flux density to the more physically
meaningful \htwo \ gas mass, one needs to employ a CO-\htwo
\ conversion factor.  While there have been numerous observational
\citep{don12,mag11,gen12,hod12,sch12,bol13,bla13,fu13,san13} and theoretical
\citep{glo11,she11a,she11b,fel11b,fel12a,nar11b,nar12a,lag12} attempts to
constrain the CO-\htwo \ conversion factor, significant uncertainty
exists in the calibration.

In \citet{nar12a}, utilizing the same simulations presented
here\footnote{In practice, in the \citet{nar11b} and \citet{nar12a}
  studies, we utilized \turtlebeach \ \citep{nar06b,nar08a} as the CO
  radiative transfer code, rather than \despotic.  However, since the
  core escape probability algorithms in \turtlebeach \ are the same as
  in \despotic, and very little non-local line coupling is expected on
  galaxy-wide scales, we expect the results to be nearly identical
  between the two.} to derive a smooth functional form for the
CO-\htwo \ conversion factor in galaxies:
\begin{eqnarray}
\label{eq:xco}
\xco = \frac{{\rm min}\left[4,6.75 \times \langle W_{\rm CO}
    \rangle^{-0.32}\right] \times 10^{20}}{Z'^{0.65}}\\
\alpha_{\rm CO} = \frac{{\rm min}\left[6.3,10.7 \times \langle W_{\rm CO}\rangle^{-0.32} \right]}{Z'^{0.65}}
\label{eq:xco2}
\end{eqnarray}
where \xco \ is in \xcounits, \alphaco \ in \alphacounits, $\langle
W_{\rm CO} \rangle$ is the resolved velocity-integrated CO intensity
in K-\kms, and $Z'$ is the metallicity of the galaxy in units of solar
metallicity.  

The coupling of the model for CO SLEDs in this work, with the
formalism for deriving \htwo \ gas masses in Equations~\ref{eq:xco} and
\ref{eq:xco2}, in
principle could allow for for the direct determination of an \htwo
\ gas mass of a high-\z \ galaxy with measurements (or assumptions) of
a CO surface brightness (at any $J$ level up to $J=9-8$), a $\Sigma_{\rm
  SFR}$, and a gas phase metallicity.  This may have implications for
derived properties of high-\z \ galaxies.

For example, galaxy baryonic gas fractions at high-\z \ are measured
to be substantially larger than in the local Universe
\citep[e.g.][]{dad10a,tac10,gea11,pop12,san13}. Correctly measuring
these and relating them to galaxy formation models in which the gas
fraction is set by a combination of accretion from the IGM, star
formation and gas outflows \citep[e.g.][]{dav12,gab13} will depend not
only on the conversion from CO $J=1- 0$ to \htwo \ gas masses
\citep[c.f.][]{nar12c}, but also a correct down-conversion from the
measured excited CO emission line to the ground state.  Systematic
trends in the excitation with galaxy physical property will, of
course, muddy comparisons between observation and theory further.

Similarly, measurements of the Kennicutt-Schmidt star formation
relation at high-\z \ are sensitively dependent on gas excitation in
galaxies.  In principle, if there is a strong relationship between gas
excitation and galaxy physical conditions, one might expect that
inferred star formation laws from high-excitation lines will be
flatter than the true underlying relation.  \citet{kru07},
\citet{nar08b} and \citet{nar08d} showed that the local SFR-$L_{\rm
  HCN}$ and SFR-$L_{\rm CO}$ relation can be strongly affected by
increasing level populations of the HCN molecule.  This work was
furthered by \citet{nar11a} who used a more rudimentary model of the
calculations presented here to find that inferred linear star
formation laws at high-\z \ potentially suffered from differential
excitation, and that the true underlying slope was steeper.  As large
surveys of galaxies at high-\z \ fill in the $\Sigma_{\rm
  SFR}-\Sigma_{\rm gas}$ relation, accurate conversions from highly
excited lines to the $J=1-0$ line will become increasingly necessary.

Finally, forthcoming (sub)millimeter and radio-wave deep fields will
determine the cosmic evolution of molecular gas masses and fractions
in galaxies \citep[e.g.][]{dec13,wal13}.  At a fixed receiver tuning, the
observations will probe increasingly high excitation lines with
redshift.  Because the line ratios depend on (as observational
parameterizations) the star formation rate surface density, and the
CO-\htwo \ conversion factor depends on CO surface brightness and gas
phase metallicity, potential biases in these deep fields may impact
the predicted measurements of the molecular gas history of the
Universe.  We will discuss these issues in a forthcoming article.

\subsection{Caveats}

There are a few notable regimes where our model may potentially fail,
which we discuss here.
\subsubsection{Gas in the Vicinity of an AGN}
  The first is when considering the gas in the vicinity of an active
  galactic nucleus (AGN).  \citet{van10b} and \citet{mei13} used
  observations with the Herschel Space Observatory to constrain the
  excitation of very high $J$ lines in nearby active galaxies Mrk 231
  and NGC 6240, respectively.  Both galaxies showed extremely high
  excitation in the high $J$ CO lines.  \citet{van10b} found that UV
  radiation from star formation could account for the excitation
  through the $J=8$ state, but that the luminosities of the higher-$J$
  lines could be explained only by invoking X-ray heating from an
  accreting supermassive black hole.  Similarly, \citet{mei13}
  attributed the high excitation of gas in NGC 6240 to CO shock
  heating.  While we do include accreting black holes in our
  hydrodynamic and dust radiative transfer calculations, we cannot
  easily resolve their impact on the thermal structure of the gas.
  Accordingly, we are unable to model the potential impact of AGN on
  the $J>10$ CO emission lines.  This said, we note that our model fit
  from Table~\ref{table:unresolved_fit} performs extremely well for
  both galaxies Mrk 231 and NGC 6240 from $J=1- 0$ through $J=9-8$,
  even without the inclusion of AGN.  It is reasonable to expect,
  then, that the heating provided by star formation, coupled to the
  high densities found in these rapidly-star-forming galaxies, are
  sufficient to explain the emission up to the $J=9-8$ line without
  the aid of additional AGN-driven heating.

\subsubsection{The Effects of Dust}
\begin{figure}
\hspace{-0.75cm}
\includegraphics[scale=0.35]{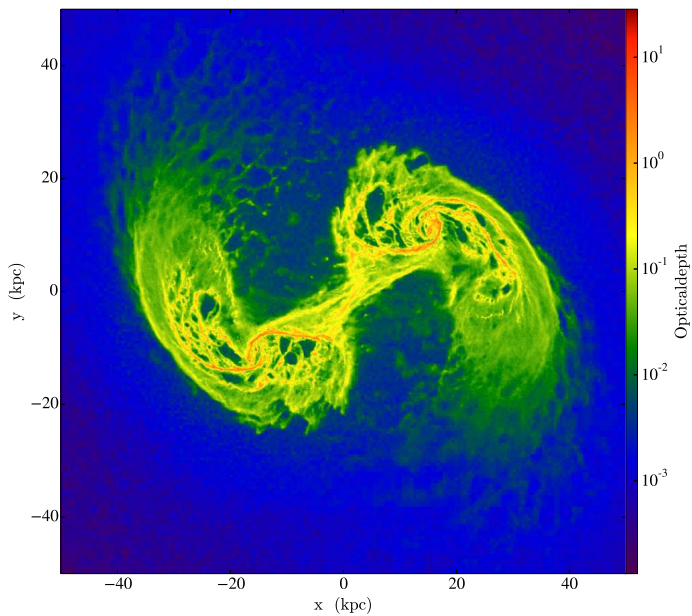}
\caption{Dust optical depth at the CO $J=9-8$ line for a simulated galaxy
  merger with $\Sigma_{\rm SFR}$ comparable to Arp 220. Opacities are
  taken from \citet{wei01}.  Even at the relatively long wavelength of
  $\sim 290$ \micron, the column densities are substantial enough that
  the lines suffer significant extinction.\label{figure:taumap}}
\end{figure}

Second, our model does not account for the potential extinction very
high $J$ CO lines may encounter after leaving their parent cloud.  A key
example of this is the extreme ULIRG, Arp 220.  Arp 220 has
potentially the highest $\Sigma_{\rm SFR}$ ever measured, at nearly
1000 \ \msunyr kpc$^{-2}$ \citep{ken98b}.  Despite this, the observed
CO SLED in this system is not comparably extreme -- rather, the high $J$
lines appear depressed compared to what one might expect given other
high $\Sigma_{\rm SFR}$ sources.  \citet{pap10c} raised the
possibility that this may owe to obscuration at FIR/submillimeter
wavelengths.  When applying a correction for this, the SLED approached
thermalized populations.  This is straightforward to understand.  The
observed gas surface densities of $\sim 2.75 \times 10^{4} \msun$
pc$^{-2}$ in Arp 220 \citep{sco97} translate to a dust optical depth of $\tau
\sim 5 $ at the wavelength of the CO $J=5-4$ line, and $\sim 15$ at
the $J=9-8$ line, assuming solar metallicities and \citet{wei01} $R_v =
3.15$ dust opacities \citep[as updated by][]{dra07}.  In
Figure~\ref{figure:taumap}, we show a projection map of the CO $J=9-8$
optical depth in a model merger with $\Sigma_{\rm SFR}$ comparable to
Arp 220.

\begin{figure}
\includegraphics[scale=0.4]{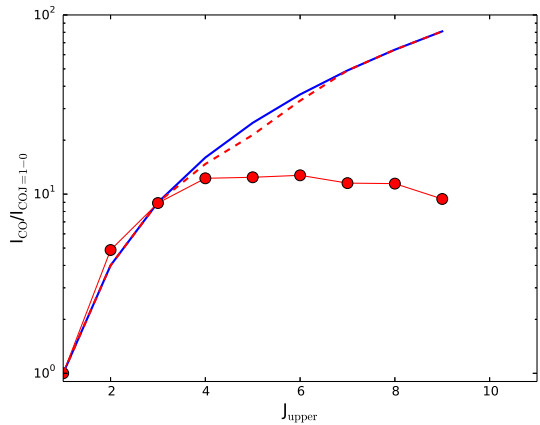}
\caption{Observed CO SLED of Arp 220 by \citet{ran11} (solid points).
  The dashed line shows our unresolved model for a galaxy of
  $\Sigma_{\rm SFR} = 1000 \ \msunyr $kpc$^{-2}$, the approximate star
  formation rate surface density of Arp 220.  The model does not
  include the effects of dust obscuration, which may be significant
  for the high $J$ lines.  See text for
  details. \label{figure:arp220}}
\end{figure}

On the other hand, \citet{ran11} utilized Herschel measurements to
derive an observed SLED for Arp 220, and found more modest dust
obscuration than the aforementioned calculation returns.  These
authors find optical depths of unity at wavelengths near the CO
$J=9-8$ line.  We show the \citet{ran11} SLED in
Figure~\ref{figure:arp220} alongside our model prediction for a
$\Sigma_{\rm SFR} \approx 1000 \ \msunyr$ kpc$^{-2}$ galaxy. Clearly
the model over-predicts the SLED compared to the \citet{ran11}
measurements.  The factor of 10 difference in estimated optical depths
from the \citet{ran11} observations, and our estimation likely drive
the difference between the observed SLED and our model SLED.  If the
\citet{ran11} dust corrections are correct, this may prove to be a
challenge for our model.  On the other hand, given the excellent
correspondence of our model with a large sample of observed galaxies
(Figure~\ref{figure:sled_multiplot}), and the extreme star formation
rate surface density of Arp 220, it may be that the dust obscuration
is in fact quite extreme in this galaxy, and the true SLED is closer
to thermalized.

\subsubsection{Excitation by the Microwave Background}

As shown in Figure~\ref{figure:physical_quantities}, the typical
minimum temperature in GMCs is $\sim 8-10$ K.  In these low density GMCs,
this simply reflects the heating expected by cosmic rays with a Milky
Way ionisation rate.  At redshifts $\z \ga 3$, however, the cosmic
microwave background (CMB) temperature will exceed this - hence
heating via this channel may be important.  Because our simulations
are not bona fide cosmological simulations, this effect thus far has
been neglected.

\begin{figure}
\hspace{-0.75cm}
\includegraphics[scale=0.5]{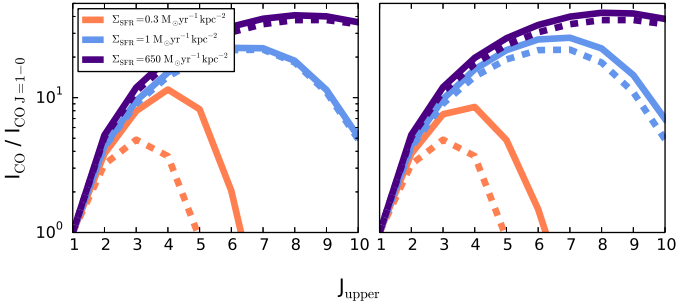}
\vspace{-3.5cm}
\caption{Effect of the cosmic microwave background (CMB) on the SLEDs
  of high-\z \ galaxies.  Left panel shows a model where $T_{\rm CMB}$
  is scaled for \z=3, while the right panel shows $T_{\rm CMB}$ for
  \z=6.  The dashed lines in each panel show our fiducial model (with
  no CMB radiation), while the solid lines show the effect of
  including the CMB.  High $\Sigma_{\rm SFR}$ galaxies are only
  marginally affected by the high temperatures of the CMB at high-\z,
  while the lowest $\Sigma_{\rm SFR}$ systems will show substantially
  more emission at high-J, and a flattening of the SLED as the CMB
  temperature increases.  Note that the effect of the CMB is only
  noticeable on galaxies that have substantially less $\Sigma_{\rm
    SFR}$ than the typical star-forming galaxy detected at \z $\ga$
  2. \label{figure:cmb}}
\end{figure}

In Figure~\ref{figure:cmb}, we show the effects of explicitly
including heating by the CMB on three simulated SLEDs for three
galaxies with a large dynamic range in $\Sigma_{\rm SFR}$. These
calculations include the effects of the CMB on both lines and dust,
following the standard treatment in \despotic\ -- see \citet{kru13}
for details. The left
panel shows a model where the CMB is inlcluded with $T_{\rm CMB} =
10.1$ K (modeling the effects at \z=3), and the right panel shows
$T_{\rm CMB} = 19.1$ K (modeling the effects at \z=6).  The dashed
  lines in each panel show our fiducial model with no CMB radiation
  included, while the solid lines show the impact of CMB variations.

The CMB has relatively minimal impact on the more intense star forming
galaxies.  This is because the typical temperatures are already quite
warm owing to an increased coupling efficiency between gas and dust,
as well as increased cosmic ray heating.  Lower $\Sigma_{\rm SFR}$
galaxies show a larger difference when considering the CMB.  The $\sim
10$ K background at \z=3 drives more excitation at lower rotational
levels (J=2), while the $\sim 20 $ K background at \z=6 spreads the
excitation to even higher levels (J=5).  The result is that at the
highest redshifts, low $\Sigma_{\rm SFR}$ galaxies will have SLEDs that
are flatter toward high-J than what our fiducial model would predict. 

This said, the effects of the CMB will be relatively minimal on any
galaxies that have been detected in CO at high-\z \ thus far.  As
shown, only galaxies with relatively low $\Sigma_{\rm SFR} \la 0.5$
\msun yr$^{-1}$ kpc$^{-2}$ show a significant impact on the CO
excitation from the CMB. This is roughly 1 order of magnitude lower
star formation rate surface density than the typical \bzk \ selected
galaxy at \zsim 2, and roughly 3 orders of magnitude lower than
typical starbursts selected at \zsim 6.  Therefore, we conclude that
the effect of CMB may drive deviations from our model in the observed
SLED for low ($\Sigma_{\rm SFR} \la 1$ \msun yr$^{-1}$ kpc$^{-2}$) SFR
surface density galaxies at high-\z, but that the effect is negligible
for present high-\z\ samples, and is likely to remain so for some time
to come.

\section{Summary and Conclusions}
\label{section:conclusions}

Observations of galaxies at low and high-\z \ show a diverse range of
CO excitation ladders, or Spectral Line Energy Distributions (SLEDs).
In order to understand the origin of variations in galaxy SLEDs, we
have combined a large suite of hydrodynamic simulations of idealized
disk galaxies and galaxy mergers with molecular line radiative
transfer calculations, and a fully self-consistent model for the
thermal structure of the molecular ISM.  Broadly, we find that the CO
ladder can be relatively well-parameterized by the galaxy-wide star
formation rate surface density ($\Sigma_{\rm SFR}$).

\begin{enumerate}
\item We find that the CO excitation increases with increasing gas
  densities and temperatures, owing to increased collisions of CO with
  \htwo \ molecules.  The average gas density and temperature in the
  molecular ISM of in turn increases with $\Sigma_{\rm SFR}$.  As a result,
  the SLED rises at higher $J$'s in galaxies with higher 
  SFR surface densities.

\item High SFR surface density galaxies tend to have increased line
  optical depths.  This effect, which comes from the increased population
  of high-$J$ levels, in turn drives down the effective density for
  thermalization of the high-$J$ lines.

\item The gas temperatures in extreme $\Sigma_{\rm SFR}$ systems rises
  owing to an increase in cosmic ray ionization rates, as well as
  increased efficiency in gas-dust energy transfer.  For J$_{\rm
    upper} \la 6$, the lines tend to be in the Rayleigh-Jeans limit
  ($E_{\rm upper} \ll kT$), and once the density is high enough to
  thermalize the levels, this produces lines whose intensity ratios
  scale as $J^2$.  Lines arising from $J\ga 7$, however, do not
  reach the Rayleigh-Jeans limit even for the
  most extreme star formation rates currently detected, and thus
  the SLED is always flatter than $J^2$ for $J \ga 7$.

\item CO line ratios can be well-parameterized by a powerlaw
  function of $\Sigma_{\rm SFR}$.  We provide the parameter fits for
  both well-resolved observations, as well as for the more typical
  unresolved observations where the same emitting area is assumed
  for all lines (Equation~\ref{eq:fit}, and
  Tables~\ref{table:resolved_fit} and \ref{table:unresolved_fit}).  Our
  functional form for the CO line ratios quantitatively matches
  multi-line observations of both local and high-\z \ galaxies over a
  large dynamic range of physical properties.  In principle, this
  model can allow for the full reconstruction of a CO SLED with an
  arbitrary CO line detection (below J$_{\rm upper} \leq 9$), and a
  star formation rate surface density estimate.

\end{enumerate}

\section*{Acknowledgements} 
DN thanks David Ballantyne, Tim Davis, Jackie Hodge, Tucker Jones,
Padelis Papadopoulos, Naseem Rangwala, Dominik Riechers, Ian Smail and
Mark Swinbank for helpful conversations, and Joe Cammisa for managing
the Haverford College Cluster, where the simulations for this project
were run.  DN also thanks Matt Turk and Nathan Goldbaum, and the
broader \yt \ community and discussion boards for helpful
conversations along the way regarding the \yt \ data analysis package.
The authours thank the anonymous referee for feedback that improved
this work.  The ideas for this work were developed at the Aspen Center
for Physics.  The Aspen Center for Physics is supported by the
National Science Foundation Grant no. PHY-1066293.  DN acknowledges
support from the NSF via grant AST-1009452. MRK acknowledges support
from: the NSF through grant AST-0955300, and NASA through ATP grant
NNX13AB84G.  This research made use of \astropy, a community-developed
core Python package for Astronomy \citep{rob13}, as well as the \yt
\ visualization and data analysis package \citep{tur11}.  This
research has made use of NASA's Astrophysics Data System Bibliographic
Services.

\end{document}